\documentclass[a4paper,fleqn,usenatbib,usedcolumn]{mnras}

\usepackage{newtxtext,newtxmath}
 
\usepackage[T1]{fontenc}
\usepackage{ae,aecompl}

\DeclareRobustCommand{\VAN}[3]{#2}
\let\VANthebibliography\thebibliography
\def\thebibliography{\DeclareRobustCommand{\VAN}[3]{##3}\VANthebibliography}

\usepackage{graphicx}	

\newcommand{\angstrom}{\mbox{\normalfont\AA}}
\defcitealias{khrykin19}{Paper~I}

\title[Dating quasars with the He\,{\normalsize\textit{II}} proximity effect]{Dating individual quasars with the He\,{\Large\textbf{II}} proximity effect}

\author[G. Worseck et al.]{
G\'abor Worseck,$^{1}$\thanks{E-mail: gworseck@uni-potsdam.de}
Ilya S. Khrykin,$^{2,3}$
Joseph F. Hennawi,$^{4,5}$
J. Xavier Prochaska,$^{6,3}$
\newauthor
and Emanuele Paolo Farina$^{7}$
\\
$^{1}$Institut f\"ur Physik und Astronomie, Universit\"at Potsdam, Karl-Liebknecht-Str.\ 24/25, D-14476 Potsdam, Germany\\
$^{2}$Southern Federal University, Stachki Avenue 194, 344090, Rostov-on-Don, Russia\\
$^{3}$Kavli Institute for the Physics and Mathematics of the Universe (WPI), UTIAS, The University of Tokyo, Kashiwa, Chiba 277-8583, Japan\\
$^{4}$Department of Physics, University of California, Santa Barbara, CA 93106, USA\\
$^{5}$Max-Planck-Institut f\"ur Astronomie, K\"onigstuhl 17, D-69117 Heidelberg, Germany\\
$^{6}$University of California - Santa Cruz, 1156 High St., Santa Cruz, CA, USA 95064\\
$^{7}$Max Planck Institut f\"ur Astrophysik, Karl--Schwarzschild--Stra{\ss}e 1, D-85748, Garching bei M\"unchen, Germany
}

\date{Accepted 2021 June 7. Received 2021 April 23; in original form 2021 January 4}

\pubyear{2021}

\begin{document}
\label{firstpage}
\pagerange{\pageref{firstpage}--\pageref{lastpage}}
\maketitle

\begin{abstract}
Constraints on the time-scales of quasar activity are key to understanding the formation
and growth of supermassive black holes (SMBHs), quasar triggering mechanisms,
and possible feedback effects on their host galaxies.
However, observational estimates of this so-called quasar lifetime are highly uncertain
($t_\mathrm{Q}\sim10^4$--$10^9$\,yr), because most methods are indirect and involve many
model-dependent assumptions.
Direct evidence of earlier activity is gained from the higher ionization state of the
intergalactic medium (IGM) in the quasar environs, observable as enhanced Ly$\alpha$
transmission in the so-called proximity zone. Due to the $\sim30$\,Myr equilibration
time-scale of \ion{He}{ii} in the $z\sim3$ IGM, the size of the \ion{He}{ii} proximity zone
depends on the time the quasar had been active before our observation
$t_\mathrm{on}\le t_\mathrm{Q}$, enabling up to $\pm0.2$\,dex precise measurements of
individual quasar on-times that are comparable to the $e$-folding time-scale
$t_\mathrm{S}\sim44$\,Myr of SMBH growth.
Here we present the first statistical sample of 13
quasars whose accurate and precise systemic redshifts allow for measurements of
sufficiently precise \ion{He}{ii} quasar proximity zone sizes between
$\simeq2$ and $\simeq15$ proper Mpc from science-grade
\textit{Hubble Space Telescope} (\textit{HST}) spectra.
Comparing these sizes to predictions from cosmological hydrodynamical simulations
post-processed with one-dimensional radiative transfer, we infer a broad range of
quasar on-times from $t_\mathrm{on}\lesssim1$\,Myr to $t_\mathrm{on}>30$\,Myr
that does not depend on quasar luminosity, black hole mass, or Eddington ratio.
These results point to episodic quasar activity over a long duty cycle, but do not
rule out substantial SMBH growth during phases of radiative inefficiency or obscuration.
\end{abstract}




\section{Introduction}

Quasars\footnote{Unless otherwise specified, we use the term ``quasar''
to refer to an unobscured accreting supermassive black hole 
that radiates at a substantial fraction of its Eddington luminosity.
}
are the most powerful sources of radiation that have emitted at an almost sustained
high luminosity during the short $\lesssim 60$\,yr time-frame accessible to modern astronomical
observations \citep{schmidt63}. Most likely they are powered by accretion of baryons onto SMBHs
\citep[e.g.][]{salpeter64,lyndenbell69,rees84}, and it is believed that past quasar phases are
required to explain the $M_\mathrm{BH}=10^9$--$10^{10}\,M_\odot$ SMBHs found in the centres
of nearby quiescent bulge-dominated galaxies \citep{soltan82,kormendy95,yu02,kormendy13}.
In numerical models of galaxy and black hole co-evolution, SMBH growth is triggered
by gas inflow from major galaxy mergers
\citep[e.g.][]{DiMatteo05,springel05a,hopkins05a,hopkins05c,hopkins06,hopkins08,capelo15,steinborn18}
and/or secular disc instabilities
\citep[e.g.][]{hopkins10,novak11,bournaud11,gabor13,hopkins16,angles-alcazar13,angles-alcazar17,angles-alcazar20},
but the physical processes on the relevant scales (sub-pc to a few pc) are still not fully understood.
In both scenarios, kinetic and thermal feedback from stars and the SMBH
self-regulate SMBH growth and obscuration. Once enough gas has been expelled,
the SMBH shines as a short-lived UV-bright quasar until feedback quenches SMBH growth,
and potentially also the star formation in the host galaxy
\citep[e.g.][]{sanders88,DiMatteo05,springel05a,hopkins05c,hopkins06}.

Although these models successfully reproduce many observed properties of galaxies
and quasars, it remains challenging for them to explain the prevalence of
$M_\mathrm{BH}=10^9$--$10^{10}\,M_\odot$ SMBHs in $z_\mathrm{em}\gtrsim 6$ quasars,
i.e.\ only $<10^9$\,yr after the Big Bang
\citep{jiang07,kurk07,mortlock11,venemans13,derosa14,wu15,mazzucchelli17,banados18,shen19,wang20,yang20}.
These early SMBHs require either quasi-continuous Eddington-limited accretion onto
massive black hole seeds \citep{sijacki09,DiMatteo12,johnson13},
super-Eddington accretion \citep{volonteri05,pacucci15,inayoshi16},
or radiatively inefficient accretion in possibly obscured phases
\citep{madau14,volonteri15,pacucci15,davies19}.

Constraining the characteristic time-scales governing quasar activity is key to understanding
the existence of early SMBHs, quasar triggering mechanisms, and whether feedback from
SMBH growth might quench black hole fuelling and star formation.
There is a growing consensus from models of galaxy and black hole co-evolution that fuelling,
feedback, and quenching are intimately intertwined, which conspire to produce episodic
quasar activity on a wide range of time-scales
\citep[$10^4$--$10^8$\,yr,][]{ciotti01,DiMatteo05,hopkins05c,hopkins06,hopkins09,hopkins16,novak11,gabor13,steinborn18,angles-alcazar17,angles-alcazar20},
often with significant variability in the accretion rate down to the time resolution limit of
the simulation \citep[10--100\,yr, e.g.][]{novak11,angles-alcazar20}.
Such short-term changes in the accretion rate may explain why some quasars show strong
variability in their luminosity and/or their emission lines on time-scales of days to decades
\citep{lamassa15,runnoe16,macleod16,mcelroy16,yang18}.
However, for longer time-scales the constraints from observations are uncertain by several
orders of magnitude \citep[e.g.][]{martini04}, because the methods
(i) are necessarily more indirect, (ii) are sensitive to particular time-scales,
(iii) often yield a population average, and (iv) involve many model-dependent assumptions.

Comparisons of the quasar number density to their host dark matter halo abundance
inferred from quasar clustering studies constrain the quasar duty cycle $t_\mathrm{dc}$,
i.e.\ the total time over the age of the Universe that a galaxy hosts a quasar \citep{haiman01,martini01}.
Due to varying assumptions on how quasars populate dark matter haloes
applied to partially discrepant quasar clustering measurements at $z_\mathrm{em}\sim 2$--4,
the inferred quasar duty cycle spans a wide range $10^6\,\mathrm{yr}\lesssim t_\mathrm{dc}\lesssim 10^9\,\mathrm{yr}$,
and may depend on redshift and/or luminosity \citep{porciani04,croom05,adelberger05,shen07,white08,white12,eftekharzadeh15}.
Alternatively, the quasar duty cycle can be estimated by extending the \citet{soltan82} argument such that 
that the quasar luminosity function traces the gas accretion history onto SMBHs
in present-day early-type galaxies. For present-day $M_\mathrm{BH}>10^9\,M_\odot$ SMBHs
that shone as quasars at their Eddington limit the inferred duty cycle is
$t_\mathrm{dc}=$(6--30)$\times 10^7$\,yr \citep{yu02,marconi04,shankar04}.
The quasar duty cycle is a population average that is insensitive to the duration
of individual quasar episodes.

The time distribution of high-accretion events, often called the episodic quasar lifetime $t_\mathrm{Q}$,
can be predicted by current numerical models or observationally estimated based on light travel time arguments.
It has been suggested that mismatches between the level of nuclear activity and the ionization conditions
of gas in and around the host galaxies imply significant nuclear variability on time-scales $t_\mathrm{Q}\sim 0.1$\,Myr
\citep{schawinski15,sartori16,schirmer16,keel17,oppenheimer18}.
However, such short quasar lifetimes cannot explain the existence of giant ($\sim 400$\,kpc) Ly$\alpha$ nebulae
around $z_\mathrm{em}\sim 2$--3 quasars \citep{cantalupo14,hennawi15,cai17,arrigoni18},
which require sustained activity for a few Myr.
Moreover, the measured equivalent widths of Ly$\alpha$ emitters, enhanced by quasar-powered fluorescence,
suggest quasar lifetimes of 1--40\,Myr depending on the emitter sample
and the quasar opening angle \citep{adelberger06,cantalupo12,trainor13,borisova16,marino18}.
However, due to geometric dilution of the quasar flux, luminous Ly$\alpha$ emitters at distances of several Mpc
are more likely to be powered intrinsically \citep{khrykin16}.

The quasar lifetime can also be constrained from the locally enhanced UV radiation field in the quasar vicinity,
the so-called proximity effect, which manifests itself as a region of enhanced IGM Ly$\alpha$ transmission
\citep[e.g.][]{bajtlik88,scott00,dallaglio08b,calverley11}.
Because the IGM reacts to a change in the photoionization rate $\Gamma$ within a finite equilibration time-scale
$t_\mathrm{eq}\approx\Gamma^{-1}$, the existence of the proximity effect implies that quasars had been emitting
continuously for $t_\mathrm{Q}\gtrsim t_\mathrm{eq}$ \citep[e.g.][]{bajtlik88}.
With an \ion{H}{i} UV background photoionization rate $\Gamma_\ion{H}{i}\simeq 10^{-12}$\,s$^{-1}$
measured in the $2\lesssim z\lesssim 5$ \ion{H}{i} Ly$\alpha$ forest \citep{becker13} one obtains a weak
lower limit $t_\mathrm{Q}\gtrsim 0.03$\,Myr for the quasar population.

During and shortly after \ion{H}{i} reionization at $z\gtrsim 5.7$, the low IGM \ion{H}{i} Ly$\alpha$ transmission
enables measurements of well-defined sizes of \ion{H}{i} proximity zones around individual quasars
\citep{fan06,carilli10,eilers17}.
The observation of any particular quasar is only sensitive to the time the quasar had been active prior to our
observation at a random point during its current luminous episode, henceforth called the on-time $t_\mathrm{on}\le t_\mathrm{Q}$.
From their very small proximity zones given their luminosity, \citet{eilers20} concluded that 5--10 per cent of all
$z_\mathrm{em}\sim 6$ quasars had recently turned on ($t_\mathrm{on}\lesssim 0.1$\,Myr),
which is also supported by a lack of extended Ly$\alpha$ emission around them \citep{farina19}.
However, because \ion{H}{i} quickly equilibrates after quasar turn-on, $z_\mathrm{em}\sim 6$
\ion{H}{i} proximity zones are insensitive to turn-on times $t_\mathrm{on}>0.1$\,Myr unless the IGM was
significantly neutral \citep{keating15,eilers17,eilers18,davies20}.

Stronger limits on the quasar lifetime may be inferred from the transverse proximity effect of a
foreground quasar in a background line of sight via the additional light travel time.
This effect has not been unambiguously detected in the $z\sim 2$--3 \ion{H}{i} Ly$\alpha$ forest
due to overdense environments around quasar hosts, quasar obscuration, and the small quasar boost
to the overall \ion{H}{i} photoionization rate
(e.g. \citealt{liske01,croft04,hennawi07,kirkman08,prochaska13}, but see \citealt{goncalves08}).
The low UV background in the post-reionization IGM increases the chance to discover the transverse proximity
effect in the \ion{H}{i} \citep[$t_\mathrm{on}>11$\,Myr, ][]{gallerani08} and the \ion{He}{ii} Ly$\alpha$
forest \citep[$t_\mathrm{on}\gtrsim 10$--25\,Myr, ][]{jakobsen03,worseck06,worseck07,schmidt17}, but frequent
non-detections can either be explained by a young age ($t_\mathrm{on}<10$\,Myr) or obscuration \citep{schmidt18}.

Direct estimates of prolonged quasar activity can be inferred from the line-of-sight \ion{He}{ii} proximity
zones of $z_\mathrm{em}\simeq 3$--4 quasars at the tail end of the \ion{He}{ii} reionization epoch,
thanks to the long \ion{He}{ii} equilibration time-scale
$t_\mathrm{eq}\approx\Gamma_\ion{He}{ii}^{-1}\simeq 30$\,Myr \citep{khrykin16}.
This is comparable to the $e$-folding time-scale of SMBH growth $t_\mathrm{S}\sim 44$\,Myr \citep{salpeter64},
and may offer unique constraints on the range of episodic quasar lifetimes in models of galaxy and black hole
co-evolution. In \citet[][hereafter \citetalias{khrykin19}]{khrykin19} we introduced a new statistical Bayesian
method to infer on-times of individual quasars from their \ion{He}{ii} proximity zones, accounting for the
degeneracy between the initial ambient IGM \ion{He}{ii} fraction and the quasar on-time that had affected
previous analyses \citep{syphers14,zheng15}.
Applying our method to six \ion{He}{ii}-transparent quasars\footnote{
\ion{He}{ii}-transparent quasars are rare quasars with sufficient flux at \ion{He}{ii} Ly$\alpha$ to secure
a science-grade (signal-to-noise ratio $\gtrsim 3$) spectrum with \textit{HST}
\citep[e.g.][]{worseck11a,syphers12,worseck19}.}
at $z_\mathrm{em}>3.6$ we inferred $0.3$\,dex
precise on-times for two quasars ($t_\mathrm{on}\simeq 0.6$ and $5.8$\,Myr, respectively), while the remaining
quasars allowed just for a joint constraint of a short $\sim 1$\,Myr on-time due to uncertainties in the quasar
systemic redshifts.

Here we build on results from \citetalias{khrykin19}, and apply a similar statistical algorithm to the sample of
seventeen $2.74<z_\mathrm{em}<3.51$ \ion{He}{ii}-transparent quasars, twelve of which have accurate and precise
systemic redshifts. In Section~\ref{sect:sample} we describe our observations and the relevant parameters of our
quasar sample. We present measurements of the \ion{He}{ii} proximity zone sizes in Section~\ref{sect:proxzones}.
In Section~\ref{sect:methods} we summarize our numerical model, before reporting on the inferred
individual quasar on-times and their relation to quasar properties in Section~\ref{sect:results}.
We discuss our results and remaining uncertainties in Section~\ref{sect:discussion},
before concluding in Section~\ref{sect:conclusions}.

We assume a flat $\Lambda$CDM cosmology with dimensionless Hubble constant $h=0.7$
($H_0=100h$\,km\,s$^{-1}$\,Mpc$^{-1}$), density parameters
$\left(\Omega_\mathrm{m},\Omega_\mathrm{b},\Omega_\Lambda\right)=\left(0.27,0.046,0.73\right)$
for total matter, baryons, and cosmological constant,
a linear dark matter power spectrum amplitude on a scale of $8h^{-1}$ comoving Mpc $\sigma_8=0.8$,
a spectral index of density perturbations $n_s=0.96$, and a helium mass fraction $Y=0.24$,
consistent with \citet{planckcollab20}.
Proper distances are quoted explicitly in proper Mpc (pMpc).

\section{Our spectroscopic data set on He\,{\small\textbf{II}} proximity zones}
\label{sect:sample}

\subsection{\textit{HST}/COS spectra of \ion{He}{ii} proximity zones}

We use the \textit{HST} UV spectra of seventeen out of twenty $z_\mathrm{em}<3.6$ \ion{He}{ii}-transparent
quasars from \citet{worseck19}, to which we refer for a detailed description of the data reduction.
Three $z_\mathrm{em}\simeq 3$ quasars (HS~1157$+$3143, SDSS~J0924$+$4852, SDSS~J1101$+$1053)
from \citet{worseck19} were excluded due to geocoronal \ion{H}{i} Ly$\alpha$ contamination of
the \ion{He}{ii} quasar proximity zone. For two $z_\mathrm{em}\simeq 2.94$ quasars
(SDSS~J0818$+$4908, SDSS~J0936$+$2927) we used only data taken in \textit{HST}'s orbital shadow
to exclude contamination of their \ion{He}{ii} proximity zones by geocoronal \ion{N}{i}
$\lambda 1200$\,\AA\ emission, as discussed in \citet{worseck19}. Likewise, geocoronal
\ion{O}{i} was excluded for four $z_\mathrm{em}\simeq 3.28$ quasars
(HE2QS~J1706$+$5904, HE2QS~J2149$-$0859, Q~0302$-$003, HE2QS~J0233$-$0149).
Although the sightline to HE2QS~J1706$+$5904 is not suitable for studying intergalactic \ion{He}{ii}
due to an optically thick \ion{H}{i} Lyman limit system at $z=0.4040$ \citep{worseck19},
its \ion{He}{ii} proximity zone is not impacted. Table~\ref{tab:sample} lists the relevant properties
of our sample.

All 17 \textit{HST} spectra were taken with the Cosmic Origins Spectrograph \citep[COS;][]{green12},
employing the G140L grating (12 spectra) or the G130M grating (5 spectra). Their resolving power
$R=\lambda/\Delta\lambda$ varies with wavelength and with the spatial position on the detector.
Table~\ref{tab:sample} lists the appropriate values in the spectral range of interest.
The spectra were rebinned to 2--3 pixels per resolution element ($R<3000$: $\simeq 0.24$\,\AA\,pixel$^{-1}$,
$10000\le R\le 15000$: $\simeq 0.04$\,\AA\,pixel$^{-1}$, $R>15000$: $\simeq 0.03$\,\AA\,pixel$^{-1}$),
yielding a signal-to-noise ratio S/N$=$3--19 in the quasar continuum immediately redward of the \ion{He}{ii}
quasar proximity zone.

After correction for Galactic extinction, the spectra were normalized by power laws
$f_\lambda\propto\lambda^{\beta}$, fitted to spectral regions without strong emission and
absorption features, but accounting for identified partial \ion{H}{i} Lyman limit breaks \citep{worseck19}.
Since our sample lacks contemporaneous and continuous spectral coverage from the rest-frame extreme UV
to the near UV, a detailed correction for cumulative \ion{H}{i} Lyman continuum attenuation is not possible.
As such, the fitted power laws
with a typical range in slope $-3\lesssim\beta\lesssim -0.5$
do not represent the intrinsic quasar spectral energy distributions (SEDs).
The typical continuum error of a few per cent does not affect our analysis. We do not account for quasar
\ion{He}{ii} Ly$\alpha$ and metal emission that is difficult to predict in detail in quasar accretion
disc models \citep[e.g.][]{syphers11a,syphers13}. Only two quasars (SDSS~J0936$+$2927 and SDSS~J2346$-$0016)
show such features, and they are sufficiently weak that they do not change our results.
Likewise, contamination of the proximity zone by low-redshift \ion{H}{i} Lyman series lines
is expected to be weak compared to the \ion{He}{ii} Ly$\alpha$ absorption.

Together with the seven $z_\mathrm{em}>3.6$ quasars discussed in \citetalias{khrykin19},
we arrive at a total sample of 24 quasars with \ion{He}{ii} proximity zones (Table~\ref{tab:sample}).

\begin{table*}
\caption{\label{tab:sample}
Our combined sample of 24 quasars with \ion{He}{ii} proximity zones,
comprising new measurements for 17 $z_\mathrm{em}<3.6$ quasars and results for
seven $z_\mathrm{em}>3.6$ quasars from \citetalias{khrykin19}.
We list the name, position, \textit{HST}/COS resolving power
and signal-to-noise ratio near \ion{He}{ii} Ly$\alpha$ in the
quasar rest frame, quasar redshift, velocity (redshift) uncertainty,
emission line and instrument for redshift measurement,
SDSS or Pan-STARRS1 $i$ band magnitude corrected for Galactic extinction,
absolute magnitude at 1450\,\AA\ rest frame,
\ion{He}{ii}-ionizing photon production rate $Q$, measured proximity zone size $R_\mathrm{pz}$,
and inferred quasar on-time $t_\mathrm{on}$ for an assumed uniform prior on the
\ion{He}{ii} fraction $0.01\le x_{\ion{He}{ii},0}\le 1$ (Section~\ref{sect:ontimes}).
}
\scriptsize
\begin{tabular}{lccD{.}{}{0}D{.}{}{2.-1}D{.}{.}{1.4}cllcD{.}{.}{2.2}cD{,}{\,\pm\,}{3,3}c}
\hline
Quasar	&R.A.	&Decl.	&\multicolumn{1}{c}{$R$}	&\mathrm{S/N^a}&\multicolumn{1}{c}{$z_\mathrm{em}$}	&$\sigma_v$	&Line	&Instrument	&$m_i$	&\multicolumn{1}{c}{$M_{1450}$}	&$\log{\frac{Q}{s^{-1}}}$	&\multicolumn{1}{c}{$R_\mathrm{pz}$}	&$t_\mathrm{on}$\\
		&(J2000)&(J2000)&		&		&															&km\,s$^{-1}$&		&			&mag	&\multicolumn{1}{c}{mag}		&							&\multicolumn{1}{c}{pMpc}				&Myr\\
\hline
HE~2347$-$4342		&$23^\mathrm{h}50^\mathrm{m}34\fs21$ &$-43\degr25\arcmin59\farcs6$	&17000	&19	&2.8852	&$44.4$	&[\ion{O}{iii}]	&FIRE		&--$^\mathrm{b}$		&-28.69^\mathrm{b}&$57.08$	&-4.77,0.15	&--\\
HE2QS~J2149$-$0859	&$21^\mathrm{h}49^\mathrm{m}27\fs77$ &$-08\degr59\arcmin03\farcs6$	&1700	&3	&3.2358	&$656$	&\ion{C}{iv}	&CAFOS		&$18.67$	&-26.83	&$56.34$	&-0.63,2.04	&$<0.46$\\
HE2QS~J1706$+$5904	&$17^\mathrm{h}06^\mathrm{m}21\fs74$ &$+59\degr04\arcmin06\farcs4$	&1700	&4	&3.2518	&$656$	&\ion{C}{iv}	&CAFOS		&$18.86$	&-26.66	&$56.27$	&-0.04,2.03	&$<0.66$\\
SDSS~J1237$+$0126	&$12^\mathrm{h}37^\mathrm{m}48\fs99$ &$+01\degr26\arcmin07\farcs0$	&2400	&4	&3.1467	&$273$	&\ion{Mg}{ii}	&TripleSpec	&$18.78$	&-26.66	&$56.27$	&1.77,0.87	&$<1.01$\\
SDSS~J2346$-$0016	&$23^\mathrm{h}46^\mathrm{m}25\fs66$ &$-00\degr16\arcmin00\farcs4$	&2600	&8	&3.5076	&$273$	&\ion{Mg}{ii}	&TripleSpec	&$17.68$	&-27.97	&$56.79$	&2.66,0.77	&$0.31^{+0.41}_{-0.21}$\\
SDSS~J0818$+$4908	&$08^\mathrm{h}18^\mathrm{m}50\fs01$ &$+49\degr08\arcmin17\farcs0$	&2200	&4	&2.9598	&$656$	&\ion{C}{iv}	&BOSS		&$18.36$	&-26.93	&$56.38$	&2.92,2.25	&$<2.42$\\
HE2QS~J0916$+$2408	&$09^\mathrm{h}16^\mathrm{m}20\fs85$ &$+24\degr08\arcmin04\farcs6$	&1800	&4	&3.4231	&$656$	&\ion{C}{iv}	&CAFOS		&$18.52$	&-27.12	&$56.45$	&3.14,1.91	&$<2.77$\\
HS~0911$+$4809		&$09^\mathrm{h}15^\mathrm{m}10\fs01$ &$+47\degr56\arcmin58\farcs8$	&10000	&6	&3.3500	&$400$	&H$\beta$		&LUCI2		&$17.77$	&-27.84	&$56.74$	&4.21,1.19	&$1.01^{+1.29}_{-0.69}$\\
HE2QS~J0233$-$0149	&$02^\mathrm{h}33^\mathrm{m}06\fs01$ &$-01\degr49\arcmin50\farcs5$	&1700	&4	&3.3115	&$656$	&\ion{C}{iv}	&CAFOS		&$18.41$	&-27.17	&$56.47$	&4.71,1.98	&$<6.24$\\
Q~1602$+$576		&$16^\mathrm{h}03^\mathrm{m}55\fs92$ &$+57\degr30\arcmin54\farcs4$	&15000	&7	&2.8608	&$273$	&\ion{Mg}{ii}	&TripleSpec	&$17.22$	&-27.99	&$56.80$	&6.10,0.97	&$1.98^{+1.41}_{-0.94}$\\
PC~0058$+$0215		&$01^\mathrm{h}00^\mathrm{m}58\fs39$ &$+02\degr31\arcmin31\farcs4$	&2200	&4	&2.8842	&$273$	&\ion{Mg}{ii}	&TripleSpec	&$18.77$	&-26.46	&$56.19$	&7.10,0.97	&$>7.24$\\
HS~1700$+$6416		&$17^\mathrm{h}01^\mathrm{m}00\fs61$ &$+64\degr12\arcmin09\farcs1$	&2100	&15	&2.7472	&$273$	&\ion{Mg}{ii}	&TripleSpec	&$15.79$	&-29.33	&$57.34$	&7.16,1.01	&$0.80^{+0.50}_{-0.35}$\\
SDSS~J0936$+$2927	&$09^\mathrm{h}36^\mathrm{m}43\fs50$ &$+29\degr27\arcmin13\farcs6$	&2200	&4	&2.9248	&$44.4$	&[\ion{O}{iii}]	&TripleSpec	&$18.06$	&-27.20	&$56.49$	&8.59,0.15	&$11.62^{+7.37}_{-4.57}$\\
HS~1024$+$1849		&$10^\mathrm{h}27^\mathrm{m}34\fs13$ &$+18\degr34\arcmin27\farcs5$	&15000	&5	&2.8521	&$273$	&\ion{Mg}{ii}	&LUCI1		&$17.66$	&-27.54	&$56.62$	&9.38,0.97	&$9.53^{+6.83}_{-4.08}$\\
SDSS~J1253$+$6817	&$12^\mathrm{h}53^\mathrm{m}53\fs71$ &$+68\degr17\arcmin14\farcs2$	&2600	&7	&3.4753	&$44.4$	&[\ion{O}{iii}]	&TripleSpec	&$18.45$	&-27.19	&$56.48$	&11.40,0.12	&$>23.55$\\
Q~0302$-$003		&$03^\mathrm{h}04^\mathrm{m}49\fs85$ &$-00\degr08\arcmin13\farcs5$	&19000	&3	&3.2850	&$44.4$	&[\ion{O}{iii}]	&TripleSpec	&$17.34$	&-28.21	&$56.89$	&13.20,0.13	&$>11.36$\\
HE2QS~J2157$+$2330	&$21^\mathrm{h}57^\mathrm{m}43\fs63$ &$+23\degr30\arcmin37\farcs3$	&1600	&4	&3.1465	&$44.4$	&[\ion{O}{iii}]	&TripleSpec	&$17.67$	&-27.77	&$56.72$	&17.40,0.14	&$>31.84$\\
\hline
HE2QS~J2354$-$2033$^\mathrm{c}$	&$23^\mathrm{h}54^\mathrm{m}52\fs00$ &$-20\degr33\arcmin20\farcs7$	&2300	&3	&3.7745	&$656$	&\ion{C}{iv}	&LRIS		&$18.90$	&-26.88	&$56.37$	&-3.65,1.68	&--\\
HE2QS~J2311$-$1417$^\mathrm{c}$	&$23^\mathrm{h}11^\mathrm{m}45\fs46$ &$-14\degr17\arcmin52\farcs1$	&2300	&4	&3.7003	&$656$	&\ion{C}{iv}	&Kast		&$18.11$	&-27.64	&$56.66$	&1.94,1.72	&$<0.86$\\
SDSS~J1614$+$4859$^\mathrm{c}$	&$16^\mathrm{h}14^\mathrm{m}26\fs81$ &$+48\degr59\arcmin58\farcs8$	&2300	&3	&3.8175	&$656$	&\ion{C}{iv}	&BOSS		&$19.45$	&-26.34	&$56.14$	&2.72,1.66	&$<5.98$\\
SDSS~J1711$+$6052$^\mathrm{c}$	&$17^\mathrm{h}11^\mathrm{m}34\fs41$ &$+60\degr52\arcmin40\farcs3$	&2700	&4	&3.8358	&$656$	&\ion{C}{iv}	&SDSS		&$19.34$	&-26.49	&$56.19$	&2.97,1.65	&$<7.48$\\
SDSS~J1319$+$5202$^\mathrm{c}$	&$13^\mathrm{h}19^\mathrm{m}14\fs20$ &$+52\degr02\arcmin00\farcs1$	&2700	&2	&3.9166	&$400$	&H$\beta$		&TripleSpec	&$17.81$	&-28.02	&$56.82$	&3.62,0.98	&$0.80^{+0.88}_{-0.50}$\\
SDSS~J1137$+$6237$^\mathrm{c}$	&$11^\mathrm{h}37^\mathrm{m}21\fs72$ &$+62\degr37\arcmin07\farcs2$	&2300	&4	&3.7886	&$656$	&\ion{C}{iv}	&BOSS		&$19.31$	&-26.46	&$56.19$	&4.92,1.68	&$>1.34$\\
HE2QS~J1630$+$0435$^\mathrm{c}$	&$16^\mathrm{h}30^\mathrm{m}56\fs34$ &$+04\degr35\arcmin59\farcs4$	&2000	&4	&3.8101	&$400$	&H$\beta$		&ISAAC		&$17.51$	&-28.37	&$56.92$	&8.43,1.02	&$2.77^{+2.27}_{-1.33}$\\
\hline
\end{tabular}
\begin{minipage}{\textwidth}
\textsuperscript{a}{Signal-to-noise ratio per pixel near \ion{He}{ii} Ly$\alpha$ ($R<3000$: $\simeq 0.24$\,\AA\,pixel$^{-1}$, $10000\le R\le 15000$: $\simeq 0.04$\,\AA\,pixel$^{-1}$, $R>15000$: $\simeq 0.03$\,\AA\,pixel$^{-1}$).}\\
\textsuperscript{b}{For HE~2347$-$4342 we obtained the observed AB magnitude $m_{1450}=16.83$ from its VLT/FORS2 spectrum calibrated to $B_\mathrm{Vega}=17.18$ \citep{worseck08}.}\\
\textsuperscript{c}{$z_\mathrm{em}>3.6$ quasar reported in \citetalias{khrykin19}. The inferred $t_\mathrm{on}$ is based on the extended grid of radiative transfer models (Section~\ref{sect:models}) and our updated definition of upper and lower limits (Section~\ref{sect:ontimes}).}
\end{minipage}
\end{table*}

\subsection{Ancillary optical and near-infrared spectra}

\subsubsection{Near-infrared spectroscopy}

We obtained near-infrared spectra of 14 quasars from our combined sample
to accurately measure their redshifts from their near-UV and optical emission lines.
We successfully observed ten quasars with the Triple Spectrograph
\citep[TripleSpec, wavelength coverage $1.0$--$2.45\,\mu$m at $R\sim 2700$;][]{herter08}
at the 200-inch Hale Telescope at Palomar Observatory in several observing campaigns between April 2011
and August 2012. Results on SDSS~J1319$+$5202 have been reported in \citetalias{khrykin19}.
The total exposure times ranged from
$0.625$ to $2.5$ hours depending on the brightness of the target and the observing conditions.
The TripleSpec spectra were reduced using a custom pipeline based on the \textsc{lowredux}
package\footnote{\href{http://www.ucolick.org/~xavier/LowRedux}{www.ucolick.org/\textasciitilde xavier/LowRedux}} \citep{prochaska09}.
Wavelength solutions were computed from OH sky emission lines, and heliocentric corrections were applied.
For all spectra, relative fluxing was performed with a telluric standard star observed close in time
and in airmass to the science target. In the TripleSpec sample the S/N per $91.4$\,km\,s$^{-1}$ pixel
ranges from 6 to 50 near the quasar emission lines of interest.

Near-infrared spectra of two further quasars were successfully obtained with the Large Binocular Telescope
Utility Cameras in the Infrared 1 and 2 \citep[LUCI1/2;][]{seifert03} in January 2014. We used the N$1.8$
cameras in seeing-limited mode. HS~1024$+$1849 was observed for a total 20 minutes with the LUCI1 G200 grating
in the wavelength range $0.95$--$1.34\,\mu$m and the $0.5\arcsec$ slit ($R\sim 2400$). HS~0911$+$4809
was observed for 20 minutes during LUCI2 commissioning with the G150 grating in the wavelength range
$2.02$--$2.28\,\mu$m and the $0.75\arcsec$ slit ($R\sim 2700$). The LUCI data were reduced with a custom set
of IRAF\footnote{IRAF \citep{tody93} is distributed by the National Optical Astronomy Observatory, which is
operated by the Association of Universities for Research in Astronomy under a cooperative agreement with the
National Science Foundation.} routines.
Each exposure was flat fielded and cleaned of cosmic ray hits using the procedure described in \citet{vandokkum01}.
Wavelength calibration was achieved by matching the position of OH sky lines in the two-dimensional spectra,
and sky subtraction was performed by subtracting subsequent pairs of frames that had been taken in a standard
ABBA dithering pattern. Relative flux calibration and telluric correction of the extracted spectra was performed
with A0V stars observed immediately after the science targets. In the wavelength range of interest the LUCI1 (LUCI2)
spectrum of HS~1024$+$1849 (HS~0911$+$4809) reaches S/N$=10$ (S/N$=5$) per $2.13$\,\AA\ ($2.67$\,\AA) pixel.

HE2QS~J1630$+$0435 from \citetalias{khrykin19} was successfully observed with the Very Large Telescope Infrared
Spectrometer And Array Camera \citep[ISAAC;][]{moorwood98}. We covered the wavelength range $2.31$--$2.43\,\mu$m
with the medium-resolution grating and the $0.6\arcsec$ slit ($R\sim 4400$) for a total exposure of $1.1$ hours
using standard AB dithering. The data were processed with the \textsc{lowredux} package. The reduced spectrum
reaches a S/N$\simeq 15$ per $\simeq 1.18$\,\AA\ pixel.

For the quasar HE~2347$-$4342 we use a reduced $R\sim 3600$ spectrum taken with the Folded-port InfraRed Echellette
\citep[FIRE;][]{simcoe08} spectrometer on the $6.5$\,m Magellan Baade Telescope, kindly provided by R.~Simcoe
(private communication). It reaches a S/N$\simeq 40$ per $12.5$\,km\,s$^{-1}$ pixel in the wavelength range of interest.

\subsubsection{Optical spectroscopy}

The remaining ten quasars from our combined sample have
only optical spectra available, which cover their rest-frame far-UV emission lines. For four quasars from
Table~\ref{tab:sample} we use $R\sim 200$ discovery spectra taken with Calar Alto Faint Object Spectrograph (CAFOS)
at the Calar Alto $2.2$\,m telescope \citep{worseck19}. HE2QS~J2311$-$1417 from \citetalias{khrykin19} had been
discovered with the Kast spectrometer at the Lick 3\,m Shane Telescope \citep{worseck19}. For HE2QS~J2354$-$2033
from \citetalias{khrykin19} we employed a higher-quality follow-up spectrum taken with the Keck Low Resolution
Imaging Spectrometer \citep[LRIS;][]{oke95,steidel04}. For the remaining 4/24 quasars from our \ion{He}{ii}
proximity zone sample that had been discovered in the Sloan Digital Sky Survey \citep[SDSS;][]{york00} we retrieved
their highest-quality spectra from SDSS Data Release~14 \citep{abolfathi18}, taken either with the original
SDSS spectrograph or the Baryon Oscillation Spectroscopic Survey (BOSS) spectrograph \citep{smee13}.

For five quasars with small \ion{He}{ii} proximity zones we also discuss below medium- and high-resolution
optical spectra of their \ion{H}{i} proximity regions.
We observed HS~0911$+$4809 with the Keck High-Resolution Echelle Spectrometer \citep[HIRES;][]{vogt94},
using the C1 decker ($0.86\arcsec$ slit, $R\sim 45,000$) and the red cross disperser. The combined reduced spectrum
reaches a S/N$\sim 60$ per $2.6$\,km\,s$^{-1}$ pixel in the \ion{H}{i} proximity region.
SDSS~J1319+5202 was observed with the Keck Echelette Spectrograph and Imager \citep[ESI;][]{sheinis02} using the
$0.75\arcsec$ slit ($R\sim 5400$), yielding a S/N$\sim 60$ per $10$\,km\,s$^{-1}$ pixel. 
The \ion{H}{i} proximity region of SDSS~J1237$+$0126 was covered in an archival Keck/HIRES spectrum \citep{omeara15}
at S/N$\sim 20$ per $2.6$\,km\,s$^{-1}$ pixel.
The VLT and Keck spectra of HE~2347$-$4342 and SDSS~J2346$-$0016 were taken from \citet{worseck16}.

\subsection{Redshift measurements}

Quasar redshifts were determined from UV-optical quasar emission lines covered in the available optical and near-infrared spectra,
accounting for their velocity shifts relative to systemic (traced by [\ion{O}{ii}] emission or \ion{Ca}{ii} absorption).
The velocity shifts are due to bulk motions in the quasar broad line region, with high-ionization lines showing
a strong dependence of the blueshift on the continuum luminosity
\citep[e.g.][]{gaskell82,tytler92,vandenberk01,richards02,richards11,hewett10,shen16a}. We measured the mode of each
emission line using the algorithm described in \citet{hennawi06}, which accounts for line blending
(in particular \ion{C}{iv}\,$\lambda1549$ blended with \ion{He}{ii}\,$\lambda1640$, and H$\beta$\,$\lambda4863$
blended with [\ion{O}{iii}]\,$\lambda\lambda4959,5007$),
and is robust against noise, line asymmetries, and associated absorption for the lines that were used eventually (S/N$>10$).
Considering our typical S/N we fitted each emission line with a single Gaussian profile.
The measured redshifts were corrected to the systemic frame by accounting for their average blueshifts,
and precisions were assigned from the standard deviations $\sigma_v$ of the velocity distributions in large
quasar samples \citep{richards02,boroson05,richards11,shen16a}. The redshift precision was our primary criterion
for the eventually adopted emission line and redshift in Table~\ref{tab:sample}:
\begin{enumerate}
\item We preferred redshifts from detected narrow [\ion{O}{iii}]\,$\lambda5007$ emission (S/N$>10$ across the line,
peak to continuum ratio $>0.3$, 5 quasars in Table~\ref{tab:sample}). We assumed a luminosity-independent
blueshift of $-27.1$\,km\,s$^{-1}$ w.r.t.\ the systemic frame
defined by low-ionization forbidden lines ([\ion{O}{ii}], [\ion{N}{ii}], [\ion{S}{ii}]), and a
velocity precision $\sigma_v=44.4$\,km\,s$^{-1}$, both based on the high-confidence sample in \citet{boroson05}.
The values are in good agreement with \citet{hewett10} and \citet{shen16a}.

\item If [\ion{O}{iii}] was not covered or not detected due to low S/N or the Baldwin effect
\citep[e.g.][]{baldwin77,stern12b,shen14,shen16b}, we took broad \ion{Mg}{ii} if the line was detected at S/N$>10$
(6 quasars).
We assumed the same luminosity-independent blueshift of $-27.1$\,km\,s$^{-1}$ as for [\ion{O}{iii}],
and a velocity precision $\sigma_v=273$\,km\,s$^{-1}$ derived by Gaussian error propagation of the
[\ion{O}{iii}] velocity precision ($44.4$\,km\,s$^{-1}$) and the \ion{Mg}{ii} velocity dispersion
w.r.t. [\ion{O}{iii}] ($269$\,km\,s$^{-1}$) from \citet{richards02}.
These values are consistent with more recent determinations \citep{hewett10,shen16b,shen16a}.

\item If neither [\ion{O}{iii}] nor \ion{Mg}{ii} was usable, but broad H$\beta$ had been covered in our $K$ band spectra,
we took the H$\beta$ line, which reasonably traces the systemic frame. We adopted a blueshift of $-109$\,km\,s$^{-1}$
and a velocity precision $\sigma_v=400$\,km\,s$^{-1}$ \citep{shen16a}.
Three quasars in our sample have H$\beta$ redshifts.

\item For the 10 quasars lacking near-infrared spectra of sufficient quality we measured the redshift from \ion{C}{iv}
covered in the optical spectra. We corrected for the known correlation of the \ion{C}{iv} blueshift with continuum
luminosity \citep{hewett10,richards11,shen16a} using a sample of lower-redshift BOSS quasar spectra covering \ion{C}{iv}
and \ion{Mg}{ii} (see \citetalias{khrykin19} for details). Our determined blueshift w.r.t. \ion{Mg}{ii}
\begin{equation}
\Delta v=\left[-192.4-599.6\log\left(\frac{1450\angstrom\times L_{\lambda,1450}}{10^{45}\mathrm{erg}\,\mathrm{s}^{-1}}\right)\right]\,\mathrm{km}\,\mathrm{s}^{-1}
\end{equation}
is in reasonable agreement with the independent determination by \citet{shen16a} for a distinct sample and line centring algorithm.
Considering the large standard deviation $\sigma_v=656$\,km\,s$^{-1}$ of the corrected \ion{C}{iv} redshifts
w.r.t.\ the \ion{Mg}{ii} redshifts, we ignored the smaller spread of the \ion{Mg}{ii} redshifts about the
systemic frame \citepalias{khrykin19}.

\end{enumerate}

For all five \ion{He}{ii}-transparent quasars with previous systemic redshift determinations
(\ion{Mg}{ii}, H$\beta$, H$\gamma$ and/or [\ion{O}{iii}]) from the literature our results are broadly consistent,
in spite of differences in the employed methods (Table~\ref{tab:zlit}). The main sources of discrepancy are
the treatment of line asymmetries and the averaging of results for multiple emission lines.

\begin{table}
\caption{\label{tab:zlit}
\ion{He}{ii}-transparent quasars with measured systemic redshifts $z_\mathrm{em}$ from this work,
which have previous redshift determinations $z_\mathrm{other}$
(\ion{Mg}{ii}, H$\beta$, H$\gamma$, [\ion{O}{iii}] or combinations thereof) from the literature.
}
\scriptsize
\begin{tabular}{lccl}
\hline
Quasar				&$z_\mathrm{em}$	&$z_\mathrm{other}$	&Reference\\
\hline
HS~1700$+$6416		&$2.7472\pm 0.0034$	&$2.7510\pm 0.0030$	&\citet{trainor12}\\
HE~2347$-$4342		&$2.8852\pm 0.0006$	&$2.887$			&\citet{syphers11b}\\
					&					&$2.886\pm 0.001$	&\citet{shull20}\\
Q~0302$-$003		&$3.2850\pm 0.0006$	&$3.2860\pm 0.0005$	&\citet{syphers14}\\
					&					&$3.2859\pm 0.0001$	&\citet{shen16b}\\
					&					&$3.287$			&\citet{zuo15}\\
					&					&$3.2896$			&\citet{coatman17}\\
SDSS~J1253$+$6817	&$3.4753\pm 0.0007$	&$3.4829$			&\citet{coatman17}\\
SDSS~J2346$-$0016	&$3.5076\pm 0.0041$	&$3.5110\pm 0.0030$	&\citet{zheng15}\\
					&					&$3.507$			&\citet{zuo15}\\
\hline
\end{tabular}
\end{table}

\subsection{Absolute magnitudes and photon production rates}

Optical photometry of the quasars was mainly obtained from SDSS Data Release 12 \citep{alam15},
with the exception of HE2QS~J2311$-$1417 that has been covered in Data Release 1 of the Panoramic Survey Telescope
and Rapid Response System 1 \citep[Pan-STARRS1;][]{chambers16,flewelling16}, and HE~2347$-$4342 that has been
taken from \citet{worseck08}. For all quasars except HE~2347$-$4342 we determined the absolute AB magnitude
at rest-frame wavelength 1450\,\AA\ from the extinction-corrected $i$ band point-spread-function AB magnitude $m_i$ as
\begin{equation}
M_{1450}\left(z_\mathrm{em}\right) = m_i-5\log\left(\frac{d_L(z_\mathrm{em})}{\mathrm{Mpc}}\right)-25-K\left(z_\mathrm{em}\right)\quad,
\end{equation}
with the luminosity distance $d_L$ in our adopted cosmology, and the bandpass correction $K$ that was obtained
by scaling the \citet{lusso15} quasar UV SED to the $i$ band flux \citep{kulkarni19}.
The latter assumption was necessary due to inaccurate relative fluxing of many of our HE2QS discovery spectra,
and we estimate an error of $0.1$\,mag for $M_{1450}$.
We used the \citet{lusso15} SED to estimate the quasar flux density at the \ion{H}{i}
Lyman limit $f_{\nu,912}$. Assuming a power-law SED $f_\nu\propto\nu^{-\alpha_\nu}$
at frequencies $\nu>\nu_{912}=3.2872\times 10^{15}$\,Hz, the total \ion{He}{ii}-ionizing photon production rate is
\begin{equation}
\label{eq:photonrate}
Q = \frac{4\pi d_L^2}{(1+z_\mathrm{em})}\int_{\nu_{228}}^\infty\frac{f_\nu}{h_\mathrm{P}\nu}\,\mathrm{d}\nu=\frac{4\pi d_L^2}{(1+z_\mathrm{em})}\frac{4^{-\alpha_\nu} f_{\nu,912}}{h_\mathrm{P}\alpha_\nu}\quad,
\end{equation}
with Planck's constant $h_\mathrm{P}$ and the \ion{He}{ii} Lyman limit frequency $\nu_{228}=4\nu_{912}$.

Due to cumulative IGM absorption by \ion{H}{i} Lyman limit systems \citep[e.g.][]{worseck11a} and the
\ion{He}{ii} Lyman series, the \ion{He}{ii}-ionizing power has to be inferred by extrapolation of the SED.
As in \citetalias{khrykin19} we assumed a power-law SED slope $\alpha_\nu=1.5$ at $\nu>\nu_{912}$,
consistent with recent measurements in stacked and composite quasar spectra \citep{shull12,stevans14,lusso15}.
Since the actual value of $\alpha_\nu$ depends on the chosen continuum windows, the total spectral coverage,
and the exclusion of weak quasar emission lines \citep{stevans14,tilton16}, the large range in $\alpha_\nu$ values
from the literature is not surprising.
SED reconstructions for two \ion{He}{ii}-transmitting quasars with complete spectral coverage indicate
quasar-to-quasar variations in $\alpha_\nu$ around our chosen value \citep{syphers13,syphers14}.
A very hard SED \citep[$\alpha_\nu=0.7$;][]{tilton16} increases $Q$ by a factor $6.5$, but may violate
constraints on the \ion{He}{ii} reionization history \citep{khaire17,kulkarni19}.
Very soft SEDs \citep[$\alpha_\nu\gtrsim 2$;][]{lusso18} are ruled out by the \textit{HST}/COS spectra,
and would lead to a modest $\sim 0.4$\,dex increase in the inferred quasar on-times \citepalias{khrykin19}.
Changes in the slope at $\nu>\nu_{228}$ do not significantly change our results \citep{khrykin16,khrykin19}.

Equation \eqref{eq:photonrate} assumes that the escape fraction of \ion{He}{ii}-ionizing photons is unity.
Recently, \citet{shull20} have suggested that many quasars could have escape fractions $\approx 0.5$--$0.9$
based on the observed range in the IGM \ion{He}{ii}/\ion{H}{i} column density ratio.
However, these observations are likely explained by radiative transfer in the IGM and the contribution
of star-forming galaxies to the \ion{H}{i}-ionizing background \citep[e.g.][]{haardt12,khaire19,puchwein19,faucher20}.
In these models the required sample-averaged \ion{H}{i} Lyman continuum escape fraction of galaxies is
$\sim 1$ per cent at $z\sim 3$, which is within recent observational constraints
\citep{grazian17,steidel18,fletcher19}. Therefore, the \ion{He}{ii}/\ion{H}{i} column density ratio
does not uniquely constrain the \ion{He}{ii}-ionizing escape fraction of unobscured quasars,
such that our assumption of a unity escape fraction is justified.

\subsection{Black hole masses and Eddington ratios}
\label{sect:bhmass}

For the subset of 14 quasars with coverage of \ion{Mg}{ii} or H$\beta$ emission in their near-infrared spectra,
single-epoch virial black hole masses were estimated from scaling relations
\begin{equation}
\label{eq:bhmass}
\log{\left(\frac{M_\mathrm{BH}}{M_\odot}\right)}=a+b\log{\left(\frac{\lambda_\mathrm{c} L_{\lambda,\lambda_\mathrm{c}}}{10^{44}\,\mathrm{erg}\,\mathrm{s}^{-1}}\right)}
+2\log{\left(\frac{\mathrm{FWHM}}{\mathrm{km}\,\mathrm{s}^{-1}}\right)}\quad,
\end{equation}
with $(\lambda_\mathrm{c},a,b)=(3000\,\angstrom,0.86,0.5)$ for \ion{Mg}{ii} \citep{vestergaard09}
and $(\lambda_\mathrm{c},a,b)=(5100\,\angstrom,0.91,0.5)$ for H$\beta$ \citep{vestergaard06}.

Due to the varying spectral coverage and quality, measurements of the full width at half maximum (FWHM)
of \ion{Mg} (H$\beta$) were made for 10 (6) quasars. We adopt the H$\beta$ measurements when available.
Table~\ref{tab:bhmass} lists the measurements and the derived quantities. The spectra were
iteratively fit with a combination of a local power-law continuum, the \citet{tsuzuki06}
\ion{Fe}{ii} emission template, and Gaussian emission line profiles. Due to the modest S/N,
a single Gaussian was considered sufficient for most \ion{Mg}{ii} lines. Each H$\beta$ emission line
was decomposed into one narrow and two broad Gaussians, and the nearby [\ion{O}{iii}]
doublet was considered simultaneously with two Gaussians for each line of the doublet. 
Bayesian joint posterior distributions of the degenerate line parameters were estimated
using the \citet{goodman10} affine-invariant ensemble sampler for Markov Chain Monte Carlo (MCMC)
as implemented in \texttt{emcee} \citep{foreman-mackey13}.
From the \ion{Mg}{ii} and H$\beta$ broad line parameter posterior distributions we computed
the posterior probability density functions (PDFs) of the total FWHM,
and adopt their median values and their equal-tailed
68 per cent credible intervals as our measurements and statistical uncertainties, respectively.
Due to the inaccurate fluxing of the near-infrared spectra, the continuum luminosities at
$\lambda_\mathrm{c}$ were estimated from $M_\mathrm{1450}$ by extrapolating the \citet{lusso15}
power-law continuum $f_\nu\propto\nu^{-0.61}$
obtained between 912\,\AA\ and 2500\,\AA, and assuming an uncertainty of $0.2$\,dex. 

The propagated statistical uncertainties in $M_\mathrm{BH}$ are much smaller than the
$\simeq 0.55$\,dex uncertainty in the scaling relations, as estimated from the
scatter of quasars with black hole masses from reverberation mapping
\citep{vestergaard06,vestergaard09,shen13}. Other scaling relations yield similar
values to within $0.1$--$0.4$\,dex \citep{mclure04,ho15,mejia-restrepo16,woo18,bahk19}.
We note that using the \citet{tsuzuki06} \ion{Fe}{ii} template instead of the \citet{vestergaard01}
template used to derive the \citet{vestergaard09} scaling relation may underestimate the black hole mass
by $\simeq 0.2$\,dex \citep{schindler20}. For the three quasars with \ion{Mg}{ii} and H$\beta$
FWHM measurements we obtained comparable virial black hole masses.

\begin{table}
\caption{\label{tab:bhmass}
Estimated black hole masses $M_\mathrm{BH}$ and Eddington ratios $L_\mathrm{bol}/L_\mathrm{Edd}$
for the 14 quasars with measured total FWHM of \ion{Mg}{ii} or H$\beta$ decomposed into $n_\mathrm{G}$
Gaussian profiles. All quoted errors are $1\sigma$ statistical uncertainties.
}
\scriptsize
\begin{tabular}{llcD{,}{\,\pm\,}{4,4}D{,}{\,\pm\,}{3,3}D{,}{\,\pm\,}{3,3}}
\hline
Quasar				&Line			&$n_\mathrm{G}$	&\multicolumn{1}{c}{FWHM}	&\multicolumn{1}{c}{$\log\left(\frac{M_\mathrm{BH}}{M_\odot}\right)$}	&\multicolumn{1}{c}{$\log\left(\frac{L_\mathrm{bol}}{L_\mathrm{Edd}}\right)$}\\
					&				&				&\multicolumn{1}{c}{km\,s$^{-1}$}	&											&\\
\hline
HE~2347$-$4342		&\ion{Mg}{ii}	&2	&4604,19	&9.44,0.10	&-0.54,0.15\\
HS~1700$+$6416		&\ion{Mg}{ii}	&2	&3837,91	&9.81,0.10	&0.05,0.15\\
HS~1024$+$1849		&\ion{Mg}{ii}	&1	&5690,415	&9.79,0.12	&-0.59,0.16\\
Q~1602$+$576		&\ion{Mg}{ii}	&1	&2285,91	&9.09,0.11	&0.28,0.15\\
PC~0058$+$0215		&\ion{Mg}{ii}	&1	&3930,320	&9.26,0.12	&-0.44,0.16\\
SDSS~J0936$+$2927	&\ion{Mg}{ii}	&1	&2650,92	&9.06,0.10	&0.02,0.15\\
SDSS~J1237$+$0126	&\ion{Mg}{ii}	&1	&2192,274	&8.79,0.15	&0.10,0.18\\
SDSS~J2346$-$0016	&\ion{Mg}{ii}	&1	&3474,137	&9.45,0.11	&-0.09,0.15\\
Q~0302$-$003		&H$\beta$		&2	&3017,320	&9.38,0.14	&0.07,0.17\\
HE2QS~J2157$+$2330	&H$\beta$		&2	&3659,324	&9.46,0.13	&-0.17,0.17\\
HS~0911$+$4809		&H$\beta$		&2	&2872,1002	&9.27,0.32	&0.05,0.34\\
SDSS~J1253$+$6817	&H$\beta$		&2	&4845,1006	&9.59,0.21	&-0.51,0.23\\
HE2QS~J1630$+$0435	&H$\beta$		&2	&4128,230	&9.69,0.11	&-0.18,0.16\\
SDSS~J1319$+$5202	&H$\beta$		&2	&5403,867	&9.85,0.17	&-0.47,0.20\\
\hline
\end{tabular}
\end{table}

Bolometric luminosities were estimated with the bolometric correction at 1450\,\AA\ from \citet{runnoe12}
\begin{equation}
\label{eq:lbol}
\log{\left(\frac{L_\mathrm{bol}}{\mathrm{erg}\,\mathrm{s}^{-1}}\right)}=4.74+0.91\log\left(\frac{1450\angstrom\times L_{\lambda,1450}}{\mathrm{erg}\,\mathrm{s}^{-1}}\right)-0.12\quad,
\end{equation}
where the last term is their advocated correction of the biased viewing angle onto the quasar accretion disc.
In a fully ionized primordial hydrogen and helium plasma the Eddington luminosity of a quasar
is \citep{shapiro05,madau14}\footnote{We note that in the literature, most expressions for the quasar Eddington luminosity incorrectly assume a pure hydrogen plasma.}
\begin{equation}
\label{eq:ledd}
L_\mathrm{Edd}=\frac{4\pi Gcm_\mathrm{p}}{\sigma_\mathrm{T}(1-Y/2)}M_\mathrm{BH}=1.43\times 10^{38}\left(\frac{M_\mathrm{BH}}{M_\odot}\right)\,\mathrm{erg}\,\mathrm{s}^{-1}\quad,
\end{equation}
where $m_\mathrm{p}$ is the proton mass, $\sigma_\mathrm{T}$ is the Thomson scattering cross section,
and natural constants are written in their usual symbols.
Expressing Equation~\eqref{eq:lbol} in magnitudes, the logarithmic Eddington ratio is
\begin{equation}
\label{eq:redd}
\log{\left(\frac{L_\mathrm{bol}}{L_\mathrm{Edd}}\right)}=-0.364 M_{1450}-\log{\left(\frac{M_\mathrm{BH}}{M_\odot}\right)}-0.82\quad.
\end{equation}
Again, the statistical uncertainties in the Eddington ratios listed in Table~\ref{tab:bhmass} are smaller than
the $\simeq 0.55$\,dex systematic uncertainty induced by the virial black hole mass scaling relations.

\section{Measured He\,{\small\textbf{II}} proximity zone sizes}
\label{sect:proxzones}

\begin{figure*}
\includegraphics[width=\textwidth]{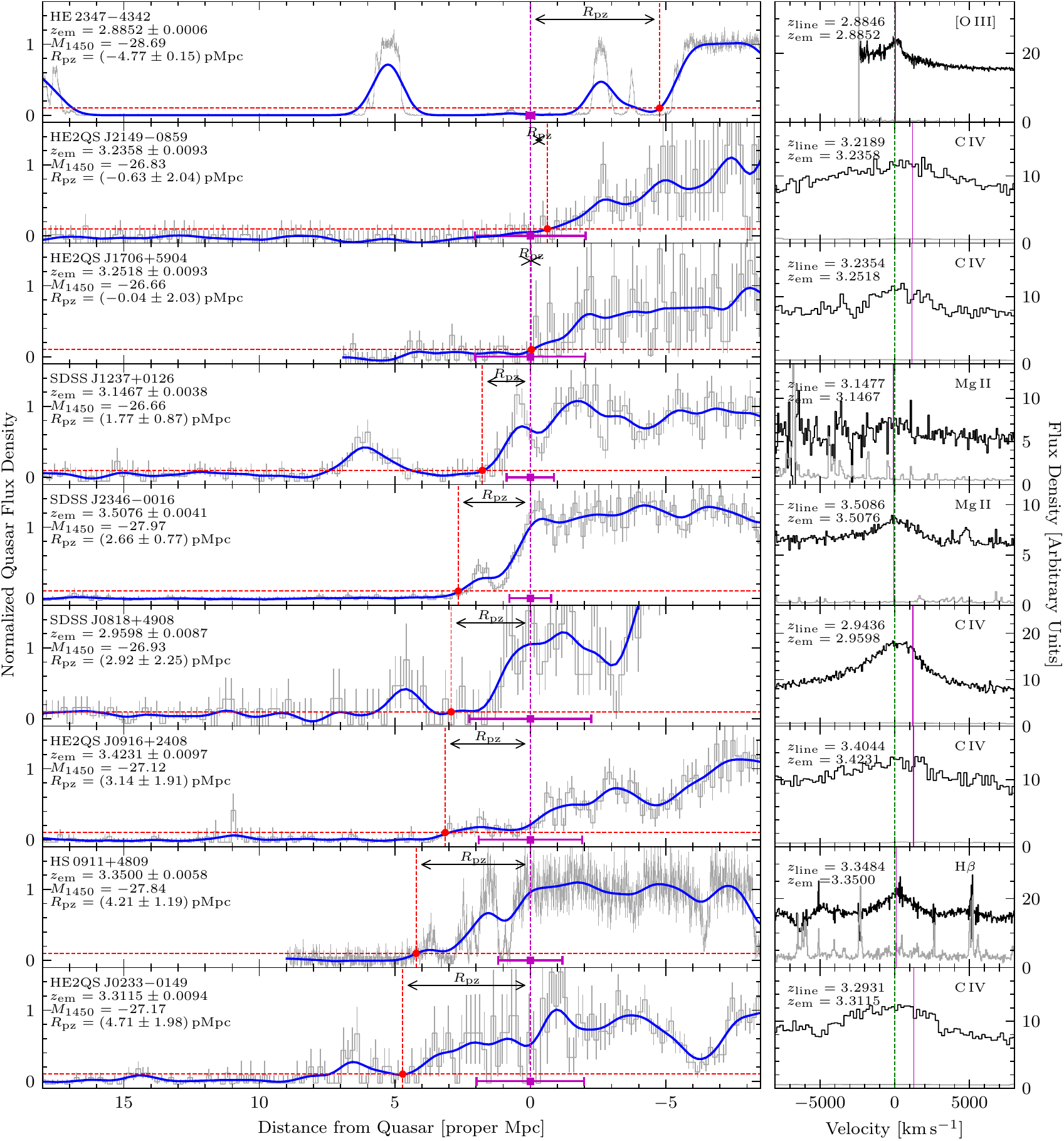}
\caption{\label{fig:he2proxspc1}
\emph{Left: }Normalized \textit{HST}/COS spectra (grey; overplotted error bars are statistical
$1\sigma$ Poisson errors) of nine $z_\mathrm{em}<3.6$ quasars from Table~\ref{tab:sample} with measured
small \ion{He}{ii} proximity zone sizes $R_\mathrm{pz}<5$\,pMpc (labelled). Distances are for the
\ion{He}{ii} Ly$\alpha$ transition relative to the quasar at the estimated systemic redshift $z_\mathrm{em}$.
With this sign convention, absorption at distances $\ll 0$ is due to infalling \ion{He}{ii}
gas clouds or foreground \ion{H}{i}, while absorption at large distances is dominated by
intergalactic \ion{He}{ii}. The violet squares with error bars mark the quasar redshift uncertainties.
The blue lines show the normalized flux smoothed with a Gaussian filter with FWHM$=1$\,pMpc.
The red dots mark the measured $R_\mathrm{pz}$, defined as the position where the smoothed flux falls below $0.1$.
\emph{Right: }Spectral regions of the labelled quasar emission lines used to estimate the systemic redshifts.
The flux density is shown in black, whereas the grey lines show the corresponding $1\sigma$ error array.
Vertical dashed lines mark the measured mode of the emission lines, yielding the emission line redshifts
$z_\mathrm{line}$. The vertical solid lines mark the applied velocity offsets to estimate the systemic
redshifts $z_\mathrm{em}$.
}
\end{figure*}

\begin{figure*}
\includegraphics[width=\textwidth]{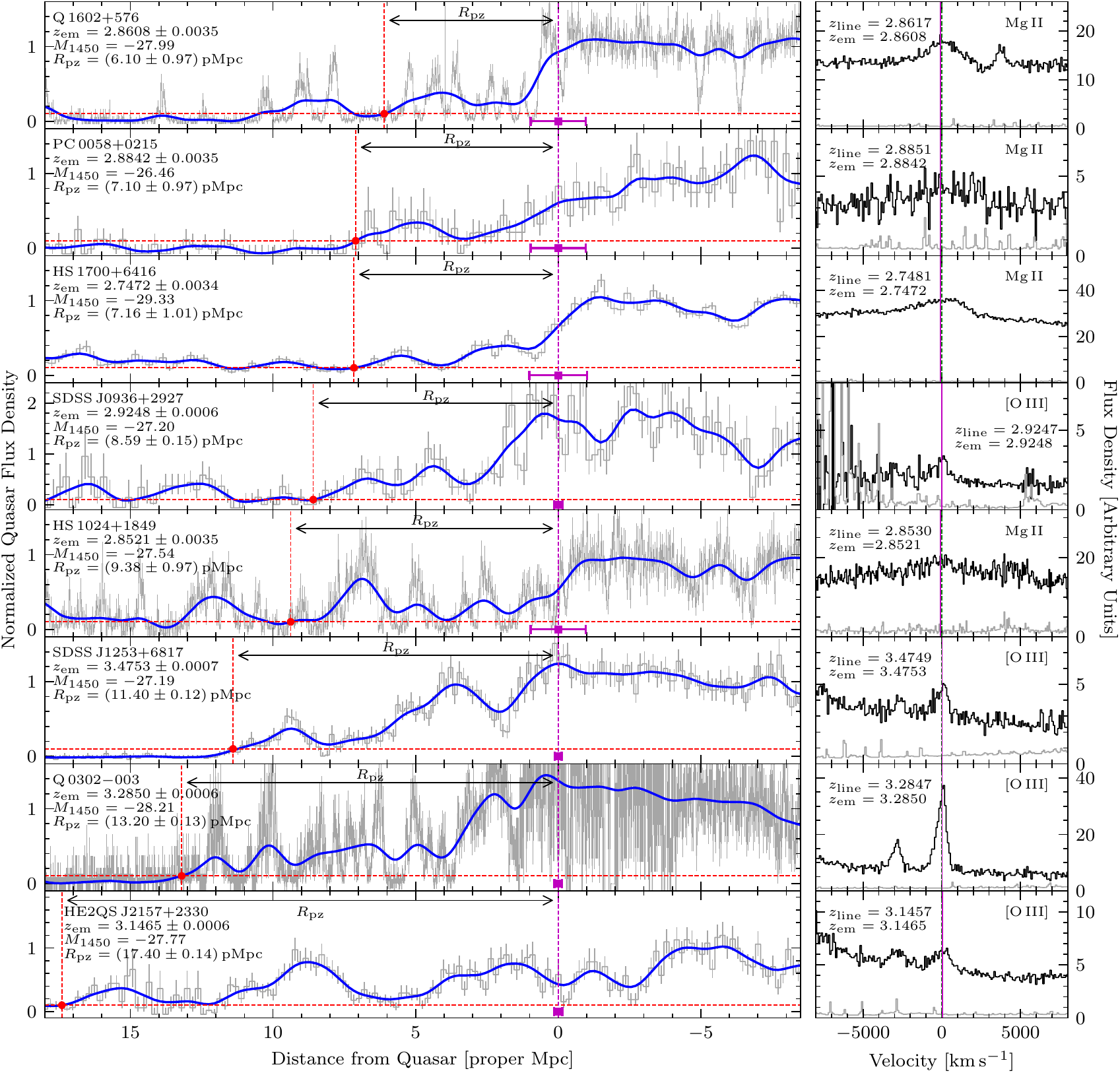}
\caption{\label{fig:he2proxspc2}
Similar to Fig.~\ref{fig:he2proxspc1} for eight $z_\mathrm{em}<3.6$ quasars with large \ion{He}{ii} proximity zone sizes
$R_\mathrm{pz}>5$\,pMpc.
}
\end{figure*}

In the quasar proximity zone the gas is more highly ionized, visible in the spectrum as excess
Ly$\alpha$ transmission. One wishes to robustly quantify its extent while limiting the impact of small-scale
density fluctuations and observational effects in heterogeneous samples, such as differences in spectral
resolution and S/N. To that aim, we adopted a procedure similar to work on \ion{H}{i} quasar proximity zones
at $z_\mathrm{em}\sim 6$ \citep[e.g.][]{fan06,carilli10,eilers17}, and smoothed the normalized
\textit{HST}/COS spectra with a Gaussian filter with an FWHM of 1\,pMpc at the respective quasar redshift.
The smoothing FWHM corresponds to $4.5$--$8.5$ pixels in our COS G140L spectra binned to $0.24$\,\AA\,pixel$^{-1}$.
The proximity zone size $R_\mathrm{pz}$ is then defined as the cosmological proper distance to the first pixel where
the smoothed normalized flux drops below $0.1$. At $z>3$ the intergalactic \ion{He}{ii} Ly$\alpha$
transmission on similar scales rarely exceeds this threshold \citep{worseck16,worseck19}, such that the \ion{He}{ii}
proximity zones are well defined. At lower redshifts the \ion{He}{ii} proximity zone sizes become less distinct
due to the emerging post-reionization \ion{He}{ii} Ly$\alpha$ forest. However, our radiative transfer simulations
(Section~\ref{sect:models}) account for density fluctuations in a predominantly ionized IGM with an initial
\ion{He}{ii} fraction as low as 1 per cent, consistent with the inferences from the \ion{He}{ii} Ly$\alpha$
forest \citep{worseck19}.
Furthermore, we verified with realistic mock spectra that our \textit{HST}/COS spectra are of sufficient quality
to yield robust proximity zone sizes (Appendix~\ref{sect:mocks}). Consequently, our quoted uncertainties
$\sigma_{R_\mathrm{pz}}$ on the proximity zone sizes are based on the individual quasar redshift uncertainties,
ranging from $\sim 0.1$\,pMpc ([\ion{O}{iii}]) to $\simeq 2$\,pMpc (\ion{C}{iv}).

\begin{figure*}
\includegraphics[width=0.99\textwidth]{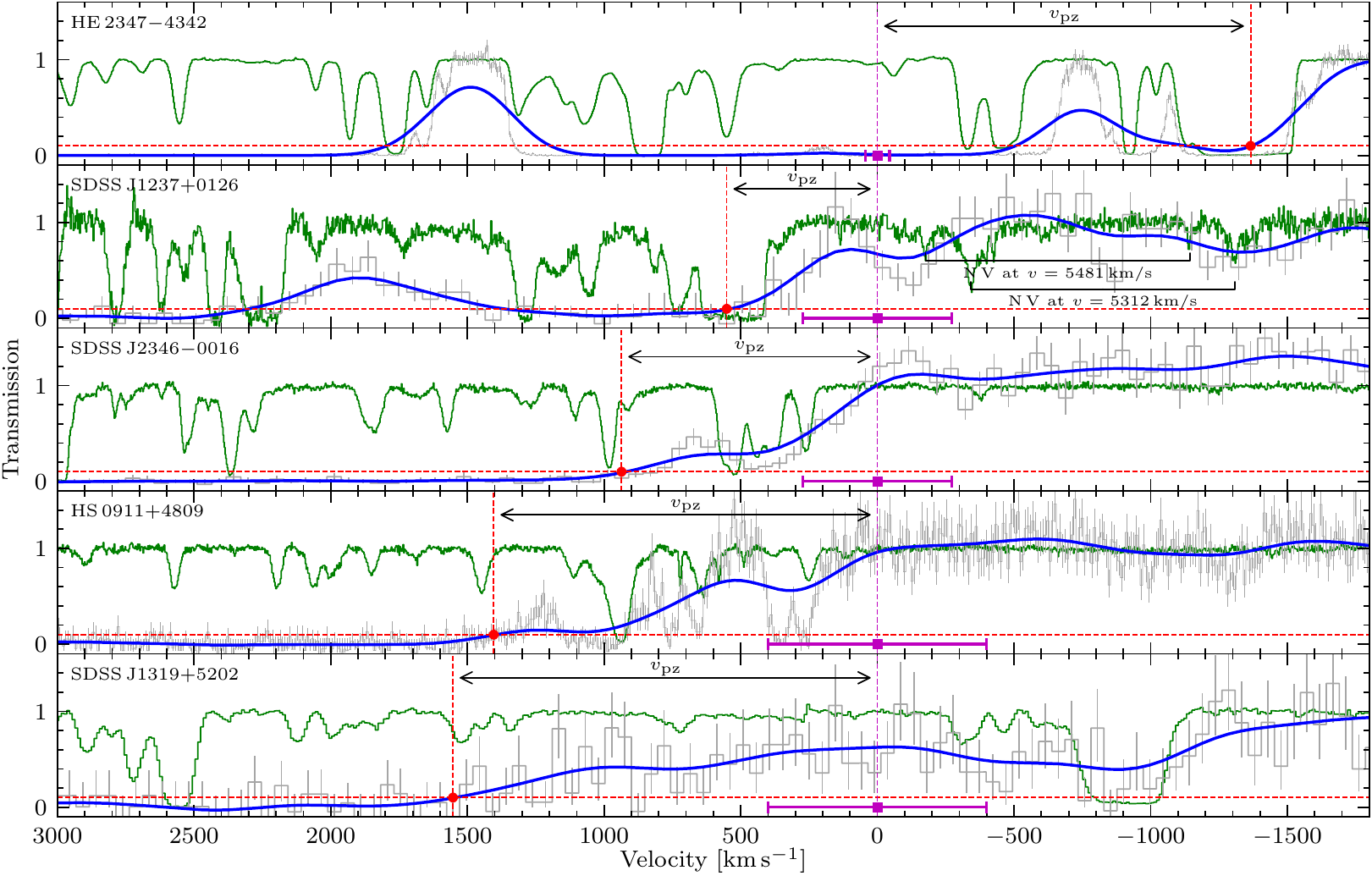}
\caption{\label{fig:he2proxassoc}
Normalized \textit{HST}/COS UV spectra (grey) and optical spectra (green) of the five quasars with sufficiently
precise redshifts ($\sigma_v\le 400$\,km\,s$^{-1}$; violet square with error bar) to measure small \ion{He}{ii}
proximity zones ($R_\mathrm{pz}<5$\,pMpc).
Velocities are for \ion{He}{ii} and \ion{H}{i} with respect to the quasar, with $v<0$ indicating infalling gas.
As in Fig.~\ref{fig:he2proxspc1} we also show the smoothed normalized UV flux (blue) with the corresponding
velocity width $v_\mathrm{pz}$ of the \ion{He}{ii} proximity zone. Note the different resolving powers for the
\ion{H}{i} spectra (SDSS~J1319$+$5202: $R\sim 5400$; $R\sim 45,000$ otherwise) and the \ion{He}{ii} spectra
(HE~2347$-$4342: $R\simeq 17,000$; HS~0911$+$4809: $R\simeq 10,000$; $R\simeq 2400$--2700 otherwise).
}
\end{figure*}

Figures~\ref{fig:he2proxspc1} and \ref{fig:he2proxspc2} show the \ion{He}{ii} proximity zone spectra of the
17 $z_\mathrm{em}<3.6$ quasars from Table~\ref{tab:sample}, ordered by their measured $R_\mathrm{pz}$.
We also show the spectral regions of the UV-optical emission lines used to determine their systemic redshifts.
The smoothed normalized flux drops significantly within several proper Mpc of each quasar, which a posteriori
justifies our chosen smoothing scale. The flux shows some contamination from unrelated low-redshift absorption
(e.g.\ at negative distances in Fig.~\ref{fig:he2proxspc1} and \ref{fig:he2proxspc2}), but this contamination
is very unlikely to affect the measurements of \ion{He}{ii} proximity zone sizes. We see strong diversity in
the proximity zone sizes and flux profiles. Most \ion{He}{ii} proximity zones show structure due to IGM density
fluctuations and radiative transfer effects, both of which have been probed with high-resolution optical spectra
of the coeval \ion{H}{i} Ly$\alpha$ absorption
\citep{reimers97,hogan97,anderson99,heap00,smette02,fechner07,shull10,syphers13,syphers14,zheng19}.

The accurate and precise redshifts of all quasars plotted in Fig.~\ref{fig:he2proxspc2} suggest that large
\ion{He}{ii} proximity zone sizes $R_\mathrm{pz}>5$\,pMpc are common. Even with less precise redshifts their
proximity zones would be much larger than the ones shown in Fig.~\ref{fig:he2proxspc1}. Measurements of small
\ion{He}{ii} proximity zones are often hampered by the large \ion{C}{iv} quasar redshift uncertainty,
limiting the further use of ten quasars from our combined sample (Table~\ref{tab:sample}).
The five $z_\mathrm{em}<3.6$ quasars with \ion{C}{iv} redshifts are marked in Fig.~\ref{fig:he2proxspc1}.
The significant absorption at negative distances in the spectra of HE2QS~J2149$-$0859, HE2QS~J1706$+$5904,
and HE2QS~J0916$+$2408 in Fig.~\ref{fig:he2proxspc1} suggests very large blueshifts of \ion{C}{iv} relative
to systemic, similar to HE2QS~J2354$-$2033 from \citetalias{khrykin19}.
In Fig.~\ref{fig:he2proxassoc} we show the \ion{He}{ii} and the corresponding \ion{H}{i} spectra of the five quasars
that have sufficiently precise redshifts to measure small \ion{He}{ii} proximity zone sizes $R_\mathrm{pz}<5$\,pMpc
($\sigma_v\le 400$\,km\,s$^{-1}$ corresponding to $\sigma_{R_\mathrm{pz}}<1.2$\,pMpc).
Coincident absorption in \ion{H}{i} and \ion{He}{ii} identifies gas flows in the quasar environment.
Large infall velocities $v\lesssim -1000$\,km\,s$^{-1}$ may not be correctly captured by our simulations,
hampering estimates of the quasar on-time.

The luminous quasar HE~2347$-$4342 lacks a \ion{He}{ii} proximity zone, which is likely explained by its peculiar
associated absorption system \citep{reimers97,fechner04}. Many of its \ion{N}{v} and \ion{O}{vi} bearing components
are consistent with being photoionized by the quasar \citep{fechner04}. Because their \ion{He}{ii} column densities
are insufficient to shield \ion{He}{ii}-ionizing photons, \citet{shull10} argued that HE~2347$-$4342 has recently
turned on ($t_\mathrm{on}<1$\,Myr). However, the location of the absorbing gas is unconstrained. 
The accurate and precise [\ion{O}{iii}] redshift $z_\mathrm{em}=2.8852\pm 0.0006$ implies large gas infall velocities
of $\sim 1500$\,km\,s$^{-1}$ onto the circumnuclear region (Fig.~\ref{fig:he2proxassoc}), and even higher velocities
if the gas resides in the circumgalactic medium of the host galaxy. Redshift space distortions \citep{hui97,weinberg97}
significantly affect the \ion{He}{ii} transmission profile, possibly masking the \ion{He}{ii} proximity effect altogether.
Conclusive evidence for or against HE~2347$-$4342 being a young quasar requires detailed numerical modelling of gas
associated with the quasar or its host galaxy, which is outside the scope of this work.
Because of the extreme peculiar gas velocities that are not represented in our simulations
(Section~\ref{sect:models}), we exclude HE~2347$-$4342 from further discussion.

\begin{figure*}
\includegraphics[width=\textwidth]{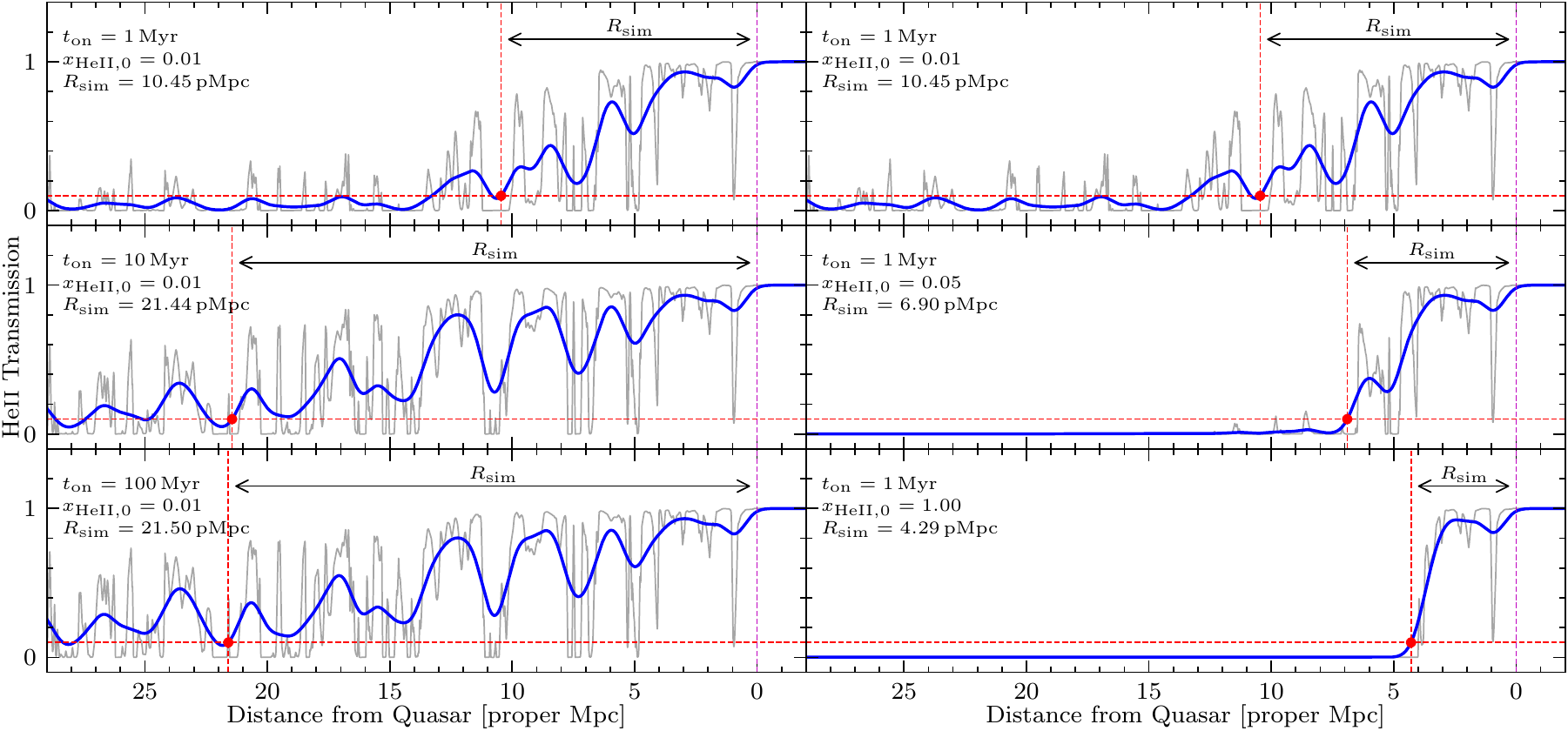}
\caption{\label{fig:q1602simspc}
Examples of \ion{He}{ii} transmission spectra from our one-dimensional radiative transfer models of Q~1602$+$576
($z_\mathrm{em}=2.8608$, $M_{1450}=-27.99$, $Q=10^{56.80}$\,s$^{-1}$). Resolved (smoothed) spectra are shown in grey (blue).
The red dots mark the proximity zone sizes $R_\mathrm{sim}$ measured from the smoothed spectra.
The left panels show the increase of $R_\mathrm{sim}$ with the quasar on-time $t_\mathrm{on}$ for a fixed initial
\ion{He}{ii} fraction $x_{\ion{He}{ii},0}=0.01$, whereas the right panels show the
decrease of $R_\mathrm{sim}$ with increasing $x_{\ion{He}{ii},0}$ for a fixed quasar on-time $t_\mathrm{on}=1$\,Myr.
}
\end{figure*}

The other four quasars shown in Fig.~\ref{fig:he2proxassoc} are less extreme. SDSS~J1237$+$0126 has a blended
optically thin \ion{H}{i} system at $v\simeq 500$\,km\,s$^{-1}$ that is likely responsible for its small measured
\ion{He}{ii} proximity zone. No high-velocity infall is detected. The \ion{He}{ii} transmission at $1500$--$2200$\,km\,s$^{-1}$
may belong to the proximity zone. However, provided that our simulations approximately capture the density and
velocity field around the quasar host halo, our strict and simple definition of $R_\mathrm{pz}$ ensures a one-to-one
comparison to our models (Section~\ref{sect:models}) that include cases of misestimated proximity zone sizes
due to the density field and peculiar velocities, similar to recent work on $z_\mathrm{em}\sim 6$ \ion{H}{i}
proximity zones \citep{eilers17,eilers20}. The strong \ion{N}{v} absorbers at $5312$ and $5481$\,km\,s$^{-1}$
may either indicate a much larger proximity zone of SDSS~J1237$+$0126 or a high-velocity outflow.
The \ion{He}{ii} spectra of SDSS~J2346$-$0016 and HS~0911$+$4809 are modulated by the density field traced by \ion{H}{i}
absorption. Their short \ion{He}{ii} proximity zones do not seem affected by peculiar velocities.
Finally, SDSS~J1319$+$5202 from \citetalias{khrykin19} shows a possibly infalling \ion{H}{i} complex at
$v\simeq -900$\,km\,s$^{-1}$ with strong associated \ion{C}{iv} and \ion{N}{v} in two components.
The complex does not affect the measurement of $R_\mathrm{pz}$ because the \ion{H}{i} absorption in the rest of
the proximity zone is much lower than expected for the $z\simeq 3.9$ \ion{H}{i} Ly$\alpha$ forest.
In summary, except for the peculiar quasar HE~2347$-$4342, gas inflows and outflows seen in absorption
do not substantially affect our measurements of small \ion{He}{ii} proximity zone sizes.
Excluding HE~2347$-$4342 and HE2QS~J2354$-$2033 from \citetalias{khrykin19} that has an anomalously large
\ion{C}{iv} blueshift, 22 quasars remain in our sample.
Thirteen of them have sufficiently precise systemic redshifts ($\sigma_v\le 400$\,km\,s$^{-1}$)
and black hole masses from near-infrared spectra.

\section{Theoretical methods}
\label{sect:methods}

\subsection{Radiative transfer models of \ion{He}{ii} proximity zones}
\label{sect:models}

To explain the diversity in the measured proximity zone sizes and to infer the on-times of the quasars,
we used a combination of hydrodynamical simulations and one-dimensional radiative transfer simulations of
\ion{He}{ii} quasar proximity zones, described in detail in \citet{khrykin16,khrykin17}.
For their cosmological setting, we used the output of a Gadget-3 \citep{springel05d} smooth particle hydrodynamics
simulation run in a cubic volume of $(25h^{-1})^3$ comoving Mpc$^3$ containing $512^3$ baryonic and
dark matter particles, respectively. Using periodic boundary conditions, we drew $1000$ one-dimensional density, velocity,
and temperature distributions (skewers) in random directions around the most massive halo in the $z_\mathrm{sim}=3.1$
snapshot of the simulation. Assuming that cosmic structure evolution is negligible between $z_\mathrm{sim}$ and the
quasar redshifts (Table~\ref{tab:sample}), we accounted for density evolution by rescaling the gas densities by
a factor $(1+z_\mathrm{em})^3/(1+z_\mathrm{sim})^3$.
The resulting skewers have a length of 160~comoving Mpc, sampled at $\mathrm{d}r=11.9$ comoving kpc,
corresponding to $\mathrm{d}v=0.86$--$0.93$~km\,s$^{-1}$ at $z_\mathrm{em}=2.74$--$3.5$.

\begin{figure*}
\includegraphics[width=0.90\textwidth]{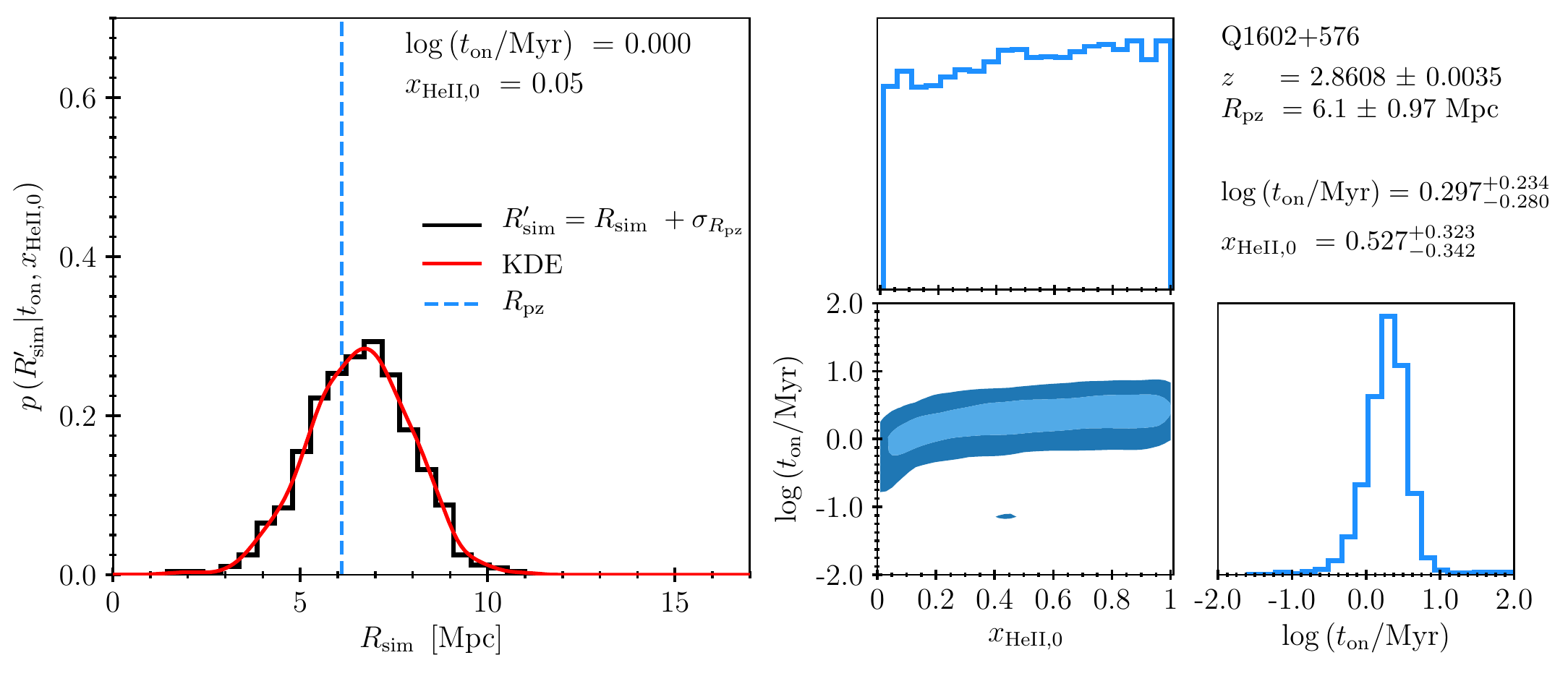}
\caption{\label{fig:kde_mcmc}
Example of our Bayesian inference of the quasar on-time for Q~1602$+$576. Left: PDF of the modelled \ion{He}{ii} proximity
zone size including measurement error as a kernel density estimate (red) of the 1000 realizations (black) for an on-time
$t_\mathrm{on}=1$\,Myr and an initial \ion{He}{ii} fraction $x_{\ion{He}{ii},0}=0.05$. The dashed line marks the measured
\ion{He}{ii} proximity zone size.
Right: Constraints on the quasar on-time and the initial \ion{He}{ii} fraction from the Bayesian inference.
The 95 per cent (dark blue) and 68 per cent per cent (light blue) credible regions from the MCMC calculations are shown.
The histograms illustrate the corresponding estimated marginalized posterior PDFs.
}
\end{figure*}

The skewers were processed with a one-dimensional radiative transfer algorithm based on the C$^2$-Ray code \citep{mellema06},
which tracks the evolution of \ion{H}{i}, \ion{He}{ii}, $e^-$, and the gas temperature to generate \ion{He}{ii} Ly$\alpha$
transmission spectra of quasar proximity zones \citep{khrykin16,khrykin17}. Analogous to \citetalias{khrykin19},
we created a set of radiative transfer models for each quasar at its respective redshift $z_\mathrm{em}$ and
photon production rate $Q$ (Table~\ref{tab:sample}), varying the quasar on-time $t_\mathrm{on}$ and the initial
\ion{He}{ii} fraction in the ambient IGM at quasar turn-on $x_{\ion{He}{ii},0}$
(or equivalently the UV background photoionization rate $\Gamma_\ion{He}{ii}$).
We assumed for simplicity that the quasars emitted continuously at their inferred luminosity for a time $t_\mathrm{on}$
prior to our observation, i.e.\ as a ``light bulb''. We considered a base-10 logarithmically spaced grid of on-times
$\log{\left(t_\mathrm{on}/\mathrm{Myr}\right)}\in\left[ -2, 2\right]$
with a step size $\Delta\log{\left(t_\mathrm{on}/\mathrm{Myr}\right)}=0.125$.
For the initial \ion{He}{ii} fraction we took the inhomogeneously spaced grid
$x_{\ion{He}{ii},0}\in\left\{0.01, 0.05, 0.1, 0.2, 0.3, 0.5, 0.6, 0.7, 0.9, 1.0 \right\}$.
Because the quasars are at lower redshifts than those from \citetalias{khrykin19} we included models with
$x_{\ion{He}{ii},0}=0.01$ representative of the IGM at the end of \ion{He}{ii} reionization \citep{khrykin16,worseck19}.
This resulted in a grid of 330 radiative transfer models per quasar, each with 1000 \ion{He}{ii} Ly$\alpha$ transmission spectra.
Redshift error is incorporated in our Bayesian inference (Section~\ref{sect:bayesinference}).

Figure~\ref{fig:q1602simspc} shows an example of model \ion{He}{ii} proximity zone spectra varying $t_\mathrm{on}$ (left)
and $x_{\ion{He}{ii},0}$ (right) for the same density skewer. The model proximity zone size $R_\mathrm{sim}$ measured
analogously to the \textit{HST}/COS spectra (Section~\ref{sect:proxzones}) depends on the quasar on-time,
as the IGM responds to changes in the radiation field on the \ion{He}{ii} equilibration time-scale
$t_\mathrm{eq}\approx\Gamma_\ion{He}{ii}^{-1}\simeq 30$\,Myr in the $z\sim 3$ IGM \citep{khrykin16}.
For $t_\mathrm{on}\lesssim t_\mathrm{eq}$ the proximity zone size increases with $t_\mathrm{on}$,
but stalls for longer on-times \citepalias{khrykin19}, as illustrated in the lower left panel
of Fig.~\ref{fig:q1602simspc}.
The proximity zone size only weakly depends on the initial IGM \ion{He}{ii} fraction due to 
the thermal proximity effect \citep{khrykin17} and the definition of the proximity zone size
that does not probe the actual size of the ionized region \citep{khrykin16}.
Only at the lowest initial \ion{He}{ii} fractions $x_{\ion{He}{ii},0}<0.05$, $R_\mathrm{sim}$ falls
at sufficiently large distances where the quasar does not dominate the total \ion{He}{ii} photoionization rate
(Fig.~\ref{fig:q1602simspc}), such that the proximity zone size becomes weakly sensitive to
$x_{\ion{He}{ii},0}$.
This weak sensitivity combined with quasar redshift uncertainties makes it challenging to infer
$x_{\ion{He}{ii},0}$ \citepalias{khrykin19}.
Furthermore, the joint distribution $R_\mathrm{sim}(t_\mathrm{on}, x_{\ion{He}{ii},0})$,
estimated from the 1000 spectra per model, is significantly blurred by density fluctuations at
low initial \ion{He}{ii} fractions. We did not account for the lower resolving power of our
\textit{HST}/COS spectra or their data quality, because they do not significantly affect
our results (Appendix~\ref{sect:mocks}).

\subsection{Bayesian inference of the quasar on-time}
\label{sect:bayesinference}

To estimate the on-times of individual quasars in our sample, we performed MCMC inference on their measured \ion{He}{ii}
proximity zones $R_\mathrm{pz}$ using the Bayesian statistical method introduced in \citetalias{khrykin19}.
First, we incorporated the individual quasar redshift uncertainties into the radiative transfer models by adding a
Gaussian-distributed random deviate with a standard deviation $\sigma_{R_\mathrm{pz}}$ (Table~\ref{tab:sample})
to each $R_\mathrm{sim}$ realization, similar to \citetalias{khrykin19}.
For each resulting distribution $R_\mathrm{sim}^\prime$ we can write a Bayesian likelihood $\mathcal{L}$ given
the combination of model parameters $\left\{t_\mathrm{on}, x_{\ion{He}{ii},0} \right\}$ per quasar,
\begin{equation}
\mathcal{L}\left( R_\mathrm{pz}|t_\mathrm{on},x_{\ion{He}{ii},0}\right) = p\left(R_\mathrm{sim}^\prime=R_\mathrm{pz}|t_\mathrm{on},x_{\ion{He}{ii},0}\right)\quad,
\label{eq:like}
\end{equation}
where $p(R_\mathrm{sim}^\prime=R_\mathrm{pz}|t_\mathrm{on},x_{\ion{He}{ii},0})$
is the PDF of the modelled \ion{He}{ii} proximity zone sizes plus redshift error $R_\mathrm{sim}^\prime$,
evaluated at the value of the measured \ion{He}{ii} proximity zone size $R_\mathrm{pz}$.
To construct this PDF we used kernel density estimation (KDE) on the respective set of 1000 model spectra.
An example KDE on the distribution of proximity zone sizes in one radiative transfer model is illustrated
in the left panel of Fig.~\ref{fig:kde_mcmc}. We then computed the likelihood of each radiative transfer model
in our $\left\{t_\mathrm{on}, x_{\ion{He}{ii},0} \right\}$ grid via Equation~\eqref{eq:like},
and constructed each quasar's continuous two-dimensional likelihood by bivariate spline interpolation.

To infer $t_\mathrm{on}$ for each individual quasar we sampled the respective likelihood with MCMC.
Because the initial \ion{He}{ii} fraction is quite uncertain at the redshifts of interest due to large-scale
UV background fluctuations at the tail end of \ion{He}{ii} reionization \citep{davies17,worseck19},
we chose to impose a uniform prior $0.01\le x_{\ion{He}{ii},0}\le 1$.
Similarly, we set a uniform prior on $\log(t_\mathrm{on}/\mathrm{Myr})$ in the range $0.01$--$100$\,Myr.
The lower limit is motivated by the ubiquitous \ion{H}{i} proximity effect implying
$t_\mathrm{on}\gtrsim\Gamma_{\ion{H}{i}}^{-1}\approx 0.03$\,Myr at $z\sim 3$ except for the few youngest
$z_\mathrm{em}\gtrsim 6$ quasars in \citet{eilers20}.
The upper limit of 100\,Myr is driven by estimates of the quasar duty cycle
\citep[$t_\mathrm{dc}=$1--1000\,Myr, e.g.][]{yu02,white08}. Furthermore, approximations of our modelling
break down at longer on-times for two reasons:
i) $t_\mathrm{on}$ becomes comparable to the cooling time meaning that cooling cannot be neglected;
ii) our assumption of a static density field for the radiative transfer in post-processing is not
valid anymore \citep{khrykin16}.

As an example, we show the results of the MCMC inference for the quasar Q~1602$+$576 in the right panels of Fig.~\ref{fig:kde_mcmc}.
Analogous to the results in \citetalias{khrykin19}, the flat posterior PDF of $x_{\ion{He}{ii},0}$ signals
the lack of sensitivity of the \ion{He}{ii} proximity zone size to the initial \ion{He}{ii} fraction in the IGM.
Yet, we are able to put tight constraints on the on-time of Q~1602$+$576 due to the small uncertainty in its
proximity zone size facilitated by its accurate and precise systemic redshift (Table~\ref{tab:sample}).

\section{Results}
\label{sect:results}

\subsection{\ion{He}{ii} proximity zone sizes do not scale with luminosity or redshift}

Many studies on $z_\mathrm{em}\sim 6$ \ion{H}{i} quasar proximity zones
\citep{maselli07,maselli09,bolton07a,lidz07,keating15,eilers17,davies20} and on $z_\mathrm{em}\sim 4$
\ion{He}{ii} proximity zones \citep{khrykin16,khrykin19} have emphasized that by definition
$R_\mathrm{pz}$ probes fully reionized gas in the quasar vicinity, and is always smaller than the actual size
of the ionized region. Moreover, simple scaling laws with luminosity $R\propto 10^{-0.4\gamma M_{1450}}$
with $\gamma=1/3$ ($\gamma=1/2$) for a neutral (ionized) IGM do not apply for smoothed spectra and
a realistic IGM density field \citep{bolton07a,davies20}.
Although the \ion{H}{i} proximity zone sizes of $z_\mathrm{em}\sim 6$ quasars generally increase with
luminosity \citep{eilers17}, 5--10 per cent of the population has undersized proximity zones that
likely indicate short on-times $t_\mathrm{on}\lesssim 0.01$--$0.1$\,Myr \citep{eilers18,eilers20}.

\begin{figure}
\includegraphics[width=\columnwidth]{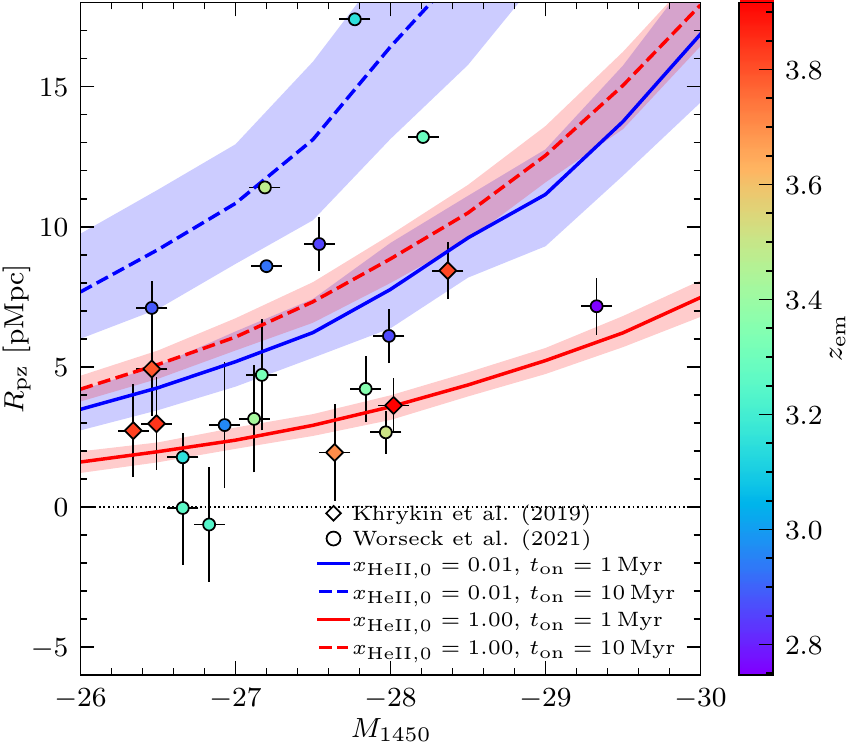}
\caption{\label{fig:rpzmag}
\ion{He}{ii} proximity zone size $R_\mathrm{pz}$ as a function of quasar absolute magnitude $M_{1450}$
for the 22 quasars in our combined sample.
Circles and diamonds mark $z_\mathrm{em}<3.6$ and higher-redshift quasars \citepalias{khrykin19}, respectively.
The colour coding indicates the quasar redshift. $R_\mathrm{pz}$ errors have been calculated from
the individual redshift errors. The dotted line marks $R_\mathrm{pz}=0$.
Overplotted is the average $R_\mathrm{pz}(M_\mathrm{1450})$ and its 16--84th percentile scatter from
our radiative transfer simulations at $z_\mathrm{em}=3.29$ for different initial
\ion{He}{ii} fractions $x_{\ion{He}{ii},0}$ and quasar on-times $t_\mathrm{on}$.
}
\end{figure}

Figure~\ref{fig:rpzmag} shows the measured \ion{He}{ii} proximity zone size $R_\mathrm{pz}$ as a function
of absolute magnitude $M_{1450}$ for the combined sample of 22 quasars, i.e.\ Table~\ref{tab:sample}
excluding HE~2347$-$4342 and HE2QS~J2354$-$2033. We see no clear relation between $R_\mathrm{pz}$ and $M_{1450}$.
For the eight $M_{1450}\sim -28$ quasars with precise systemic redshifts there is a factor $\sim 5$ spread
in $R_\mathrm{pz}$. We overplot predictions of the average $R_\mathrm{pz}(M_{1450})$ from our
radiative transfer simulations at the median redshift $z_\mathrm{em}=3.29$ of our combined
sample\footnote{We verified that these relations do not change dramatically due to IGM density evolution
in the redshift range of interest.}.
The model predictions for extreme values of the IGM \ion{He}{ii} fraction before quasar turn-on
$x_{\ion{He}{ii},0}\in\{0.01,1\}$ and two representative on-times $t_\mathrm{on}\in\{1\,\mathrm{Myr},10\,\mathrm{Myr}\}$
cover a similar range of $R_\mathrm{pz}$ as the observations.
IGM density fluctuations give rise to intrinsic scatter in the simulated $R_\mathrm{pz}$ that
increases toward low \ion{He}{ii} fractions, i.e.\ in the emerging \ion{He}{ii} Ly$\alpha$ forest.
The \ion{He}{ii} fraction and the on-time are often degenerate \citep{khrykin16,khrykin19},
as illustrated by the overlapping curves for
$(x_{\ion{He}{ii},0},t_\mathrm{on})=(0.01,1\,\mathrm{Myr})$ and $(x_{\ion{He}{ii},0},t_\mathrm{on})=(1,10\,\mathrm{Myr})$. 
Nevertheless, the smallest precisely measured \ion{He}{ii} proximity zone sizes require short on-times of $\lesssim 1$\,Myr,
whereas the largest ones require long on-times of $\gtrsim 10$\,Myr. Therefore, our measurements are sensitive
to $t_\mathrm{on}$ up to the \ion{He}{ii} equilibration time-scale $t_\mathrm{eq}\simeq 30$\,Myr at $z\sim 3$
\citep{khrykin16,khrykin19}.

Our finding that \ion{He}{ii} proximity zone sizes do not scale with luminosity unlike their $z_\mathrm{em}\sim 6$
\ion{H}{i} counterparts, can be explained by the lack of sensitivity of \ion{H}{i} proximity zone sizes to
$t_\mathrm{on}\gtrsim 0.1$\,Myr, which is rooted in the shorter \ion{H}{i} equilibration time-scale.
If most $z_\mathrm{em}\sim 6$ quasars shone longer than $0.1$\,Myr, their similar \ion{H}{i} proximity zone sizes
result in an overall strong scaling with luminosity \citep{davies20}, with shorter on-times being apparent as outliers
\citep{eilers17,eilers18,eilers20}. In contrast, \ion{He}{ii} proximity zone sizes probe $t_\mathrm{on}\lesssim 30$\,Myr,
so variations in the individual on-times can effectively remove any correlation with luminosity.

\begin{figure}
\includegraphics[width=\columnwidth]{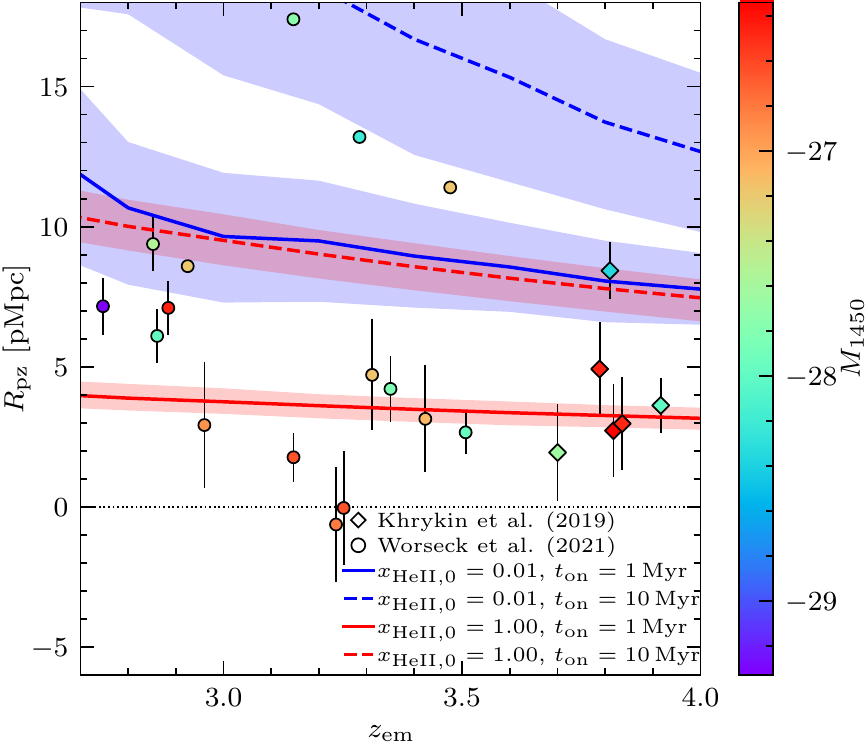}
\caption{\label{fig:rpzzem}
Similar to Fig.~\ref{fig:rpzmag} for the \ion{He}{ii} proximity zone size $R_\mathrm{pz}$
as a function of quasar redshift $z_\mathrm{em}$
The colour coding indicates the quasar absolute magnitude.
Overplotted is the average $R_\mathrm{pz}(z_\mathrm{em})$ and its 16--84th percentile scatter from
our radiative transfer simulations at $M_\mathrm{1450}=-27$ for different initial
\ion{He}{ii} fractions $x_{\ion{He}{ii},0}$ and quasar on-times $t_\mathrm{on}$.
}
\end{figure}

As shown in Fig.~\ref{fig:rpzzem} the \ion{He}{ii} proximity zone size does not depend on redshift.
Some of the scatter in $R_\mathrm{pz}$ vs.\ redshift is due to the $\simeq 3$\,mag range in $M_{1450}$,
but due to the lack of a clear correlation with luminosity we do not take out this dependence.
The large scatter in $R_\mathrm{pz}$ for the eight quasars at $M_{1450}\sim -28$ with precise
systemic redshifts confirms that there is no evidence for a scaling with redshift.
Our measurements do not support previous claims of a significant decline of the luminosity-normalized
$R_\mathrm{pz}$ at $z_\mathrm{em}>3.3$ by \citet{zheng15} for a slightly different definition of the
proximity zone size that does not track the quasar ionization front \citep{khrykin16}.
At fixed luminosity our radiative transfer simulations show a shallow decrease of $R_\mathrm{pz}$
with redshift for a large range of on-times and IGM \ion{He}{ii} fractions (Fig.~\ref{fig:rpzzem}).
This is a direct consequence of the observational definition of $R_\mathrm{pz}$ and its insensitivity
to the IGM \ion{He}{ii} fraction \citep{khrykin16,khrykin19}. Only for long on-times and small
\ion{He}{ii} fractions is there a significant decrease with redshift that is driven by IGM density evolution.
Similar shallow relations have been obtained for \ion{H}{i} proximity zones at $z_\mathrm{em}\sim 6$
\citep{eilers17,davies20}. For the model relations in Fig.~\ref{fig:rpzzem} we chose $M_{1450}=-27$,
somewhat fainter than the median $M_{1450}$ of our sample, in order to facilitate comparison with \citet{eilers17}.
For the three $M_{1450}\sim -27$ quasars from \citetalias{khrykin19} small but uncertain proximity zone sizes
indicate $t_\mathrm{on}\ll 10$\,Myr irrespective of the initial IGM \ion{He}{ii} fraction, consistent with our
joint analysis in \citetalias{khrykin19}. On the other hand, the large \ion{He}{ii} proximity zone of the
$M_{1450}\sim -27$ quasar SDSS~J1253$+$6817 at $z_\mathrm{em}=3.4753$ requires $t_\mathrm{on}>1$\,Myr
irrespective of the \ion{He}{ii} fraction. This again highlights the sensitivity of our measured \ion{He}{ii}
proximity zone sizes to the quasar on-time.

\subsection{Constraints on the quasar on-time}
\label{sect:ontimes}

\begin{figure}
\includegraphics[width=\columnwidth]{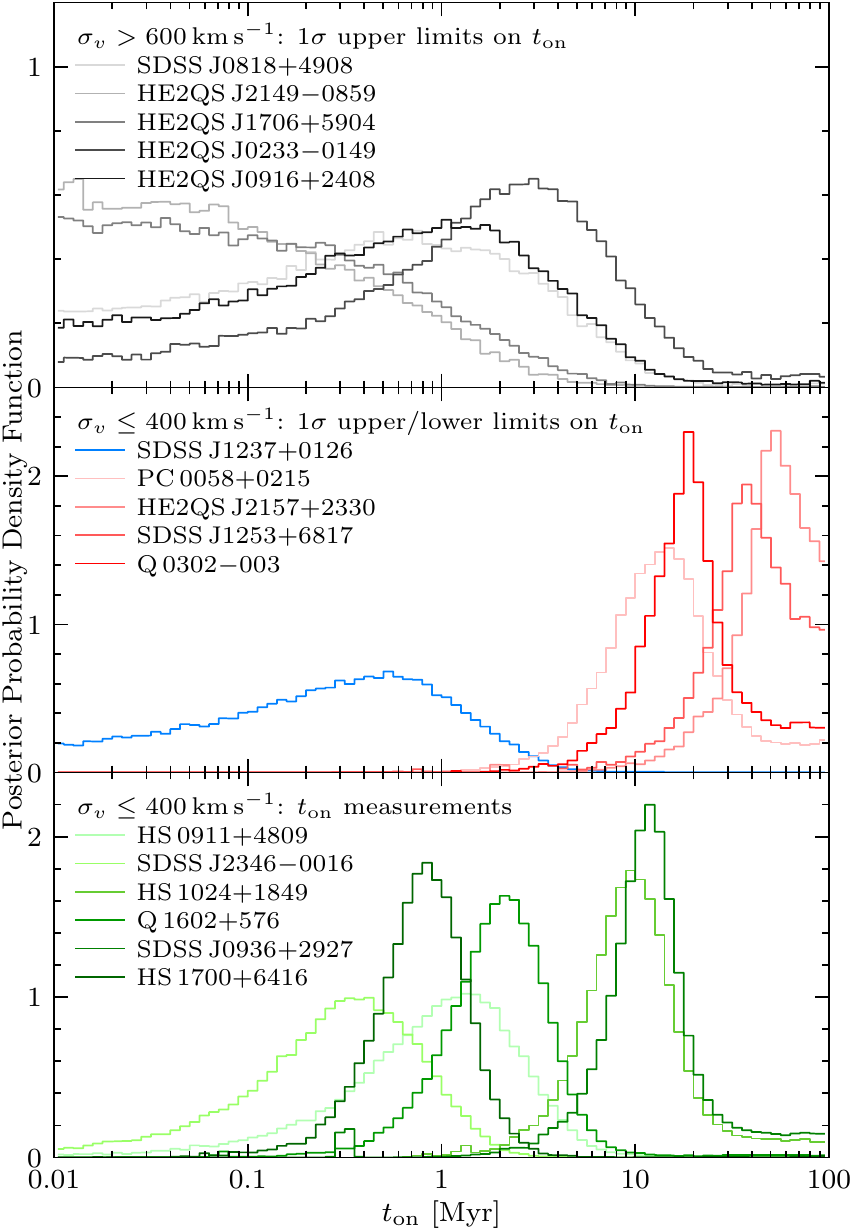}
\caption{\label{fig:he2tqpost}
MCMC estimates of the posterior PDFs of the quasar on-time $t_\mathrm{on}$ marginalized over the initial
\ion{He}{ii} fraction for the 16 newly reported $z_\mathrm{em}<3.6$ quasars (Table~\ref{tab:sample}).
The posteriors are grouped by their shape and by quasar redshift error that determine the sensitivity of
$R_\mathrm{pz}$ to $t_\mathrm{on}$. The colour coding distinguishes $t_\mathrm{on}$ posteriors for quasars
with uncertain \ion{C}{iv} redshifts (grey tones), and more precise redshifts yielding upper limits (blue),
lower limits (red tones) and measurements (green tones) for $t_\mathrm{on}$, respectively.
}
\end{figure}

Figure~\ref{fig:he2tqpost} shows the MCMC estimates of the quasar on-time posterior PDFs marginalized over
the initial \ion{He}{ii} fraction for the 16 remaining $z_\mathrm{em}<3.6$ quasars excluding HE~2347$-$4342.
Most PDFs are based on $\sim 130,000$ posterior samples depending on the width of the PDF.
Many PDFs are $\gg 0$ at the upper or lower limit of our flat logarithmic prior on the quasar on-time
$0.01$\,Myr\,$\le t_\mathrm{on}\le 100$\,Myr. Because the lower limit of our prior is physically motivated by
the \ion{H}{i} proximity effect, while the flatness of the posterior at $t_\mathrm{on}\gtrsim 30$\,Myr is
effectively determined by the equilibration time-scale, some of these posterior PDFs provide only limited
constraints on $t_\mathrm{on}$. Specifically, if the posterior at $t_\mathrm{on}=0.01$\,Myr is $>10$ per cent of its maximum,
we quote a $1\sigma$ upper limit on $t_\mathrm{on}$ as the 84th percentile of the posterior. Likewise, if the posterior at
$t_\mathrm{on}=100$\,Myr is $>10$ per cent of its maximum, we define a $1\sigma$ lower limit on $t_\mathrm{on}$ as the
16th percentile of the posterior. For the remaining quasars we quote the median of the posterior as a measurement of
$t_\mathrm{on}$ with a $1\sigma$ equal-tailed credibility interval derived from the 16th and the 84th percentile of
the posterior, respectively. 

The upper panel of Fig.~\ref{fig:he2tqpost} shows the $t_\mathrm{on}$ posteriors of the five quasars from
Table~\ref{tab:sample} whose $R_\mathrm{pz}$ values are highly uncertain due to \ion{C}{iv} redshift errors.
Large redshift errors significantly broaden the model $R_\mathrm{sim}$ distributions and result in weak constraints
on $t_\mathrm{on}$, similar to four quasars from \citetalias{khrykin19}. According to the above definition,
we obtain $1\sigma$ upper limits on $t_\mathrm{on}$. Quasars with more precise redshifts ($\sigma_v\le 400$\,km\,s$^{-1}$)
have significantly narrower $t_\mathrm{on}$ posteriors, except SDSS~J1237$+$0126 whose \ion{He}{ii} proximity zone
size may have been underestimated (Section~\ref{sect:proxzones}). For four quasars we obtain lower limits on the quasar
on-time (middle panel of Fig.~\ref{fig:he2tqpost}) due to their combination of a large proximity zone and moderate
to low luminosity (PC~0058$+$0215). The bottom panel in Fig.~\ref{fig:he2tqpost} shows the $t_\mathrm{on}$ posteriors
of the six quasars for which we obtain individual on-time measurements.
Their posteriors span distinct ranges in $t_\mathrm{on}$, indicating an intrinsically broad distribution of
quasar on-times from $\lesssim 1$\,Myr to $\sim 10$\,Myr. The precision of the measurements ranges from
$0.28$\,dex (HS~1700$+$6416) to $0.48$\,dex (HS~0911$+$4809), which mainly reflects the redshift precision.
On-times of up $\simeq 10$\,Myr can be measured securely, but for longer on-times the flat tail due to the finite
equilibration timescale becomes stronger, eventually resulting in lower limits on the on-time.
Both the middle and lower panels show that with sufficiently precise redshifts, \ion{He}{ii} proximity zones
constrain individual on-times from $\simeq 1$\,Myr up to the \ion{He}{ii} equilibration time of $\sim 30$\,Myr,
which sets the physical limit of the method.

\begin{figure}
\includegraphics[width=\columnwidth]{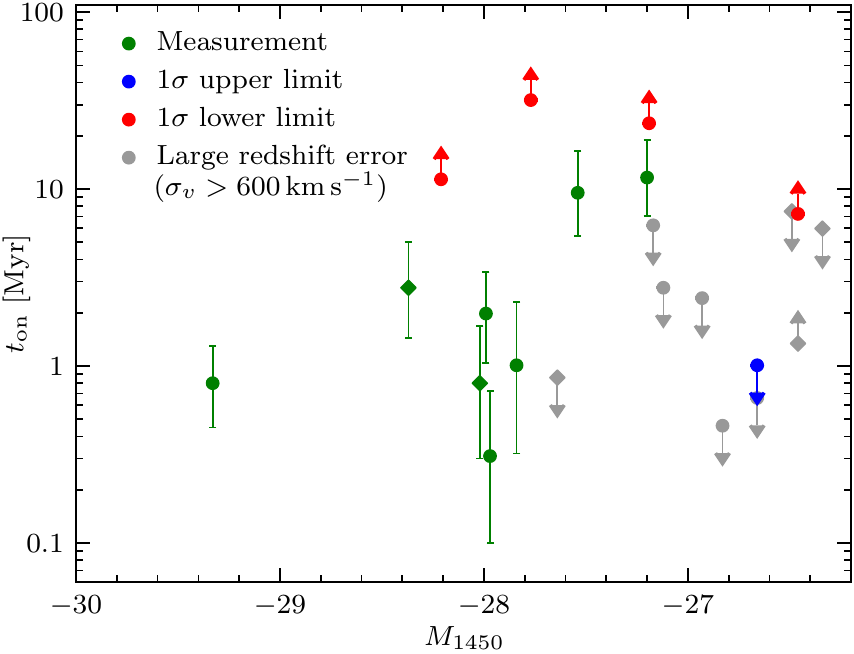}
\caption{\label{fig:he2tqmag}
Inferred quasar on-time $t_\mathrm{on}$ as a function of absolute magnitude $M_\mathrm{1450}$
for the 22 quasars in our combined sample.
Green symbols show $t_\mathrm{on}$ measurements (posterior median with 16--84th percentile range from
Fig.~\ref{fig:he2tqpost}), while blue and red symbols mark $1\sigma$ upper limits
(84th percentile of the posterior) and $1\sigma$ lower limits (16th percentile of the posterior)
for quasars with precise systemic redshifts ($\sigma_v\le 400$\,km\,s$^{-1}$), respectively.
Limits derived for quasars with larger redshift errors are shown in grey.
}
\end{figure}

In Fig.~\ref{fig:he2tqmag} we plot the inferred quasar on-times or limits thereof as a function of
quasar absolute magnitude. We also show the results for the six $z_\mathrm{em}>3.6$ quasars from
\citetalias{khrykin19} with updated inferences based on our extended model grid and adjusted to our different
definition of limits on $t_\mathrm{on}$. There is strong diversity in the quasar on-times with no obvious
dependence on absolute magnitude. This is similar to the lack of a trend in $R_\mathrm{pz}(M_{1450})$ in
Fig.~\ref{fig:rpzmag}, but now we account for the redshift dependence and provide quantitative constraints
on $t_\mathrm{on}$ based on the observed scatter of \ion{He}{ii} proximity zone sizes.

\subsection{No dependence of quasar on-time on black hole mass and Eddington ratio}

\begin{figure}
\includegraphics[width=\columnwidth]{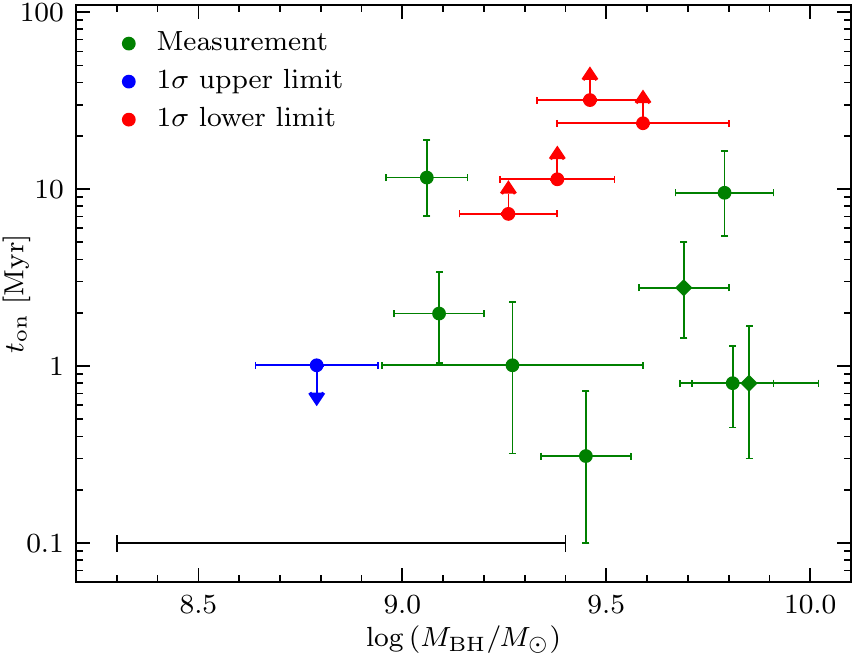}
\caption{\label{fig:he2tqmbh}
Inferred quasar on-time $t_\mathrm{on}$ as a function of black hole mass $M_\mathrm{BH}$ for
13 \ion{He}{ii}-transparent quasars (Table~\ref{tab:bhmass} excluding HE~2347$-$4342).
Quasar on-time measurements and limits are labelled as in Fig.~\ref{fig:he2tqmag}.
Individual error bars are $1\sigma$ statistical errors, while the bottom bar indicates
the $\pm 0.55$\,dex systematic uncertainty in $M_\mathrm{BH}$.
}
\end{figure}

\begin{figure}
\includegraphics[width=\columnwidth]{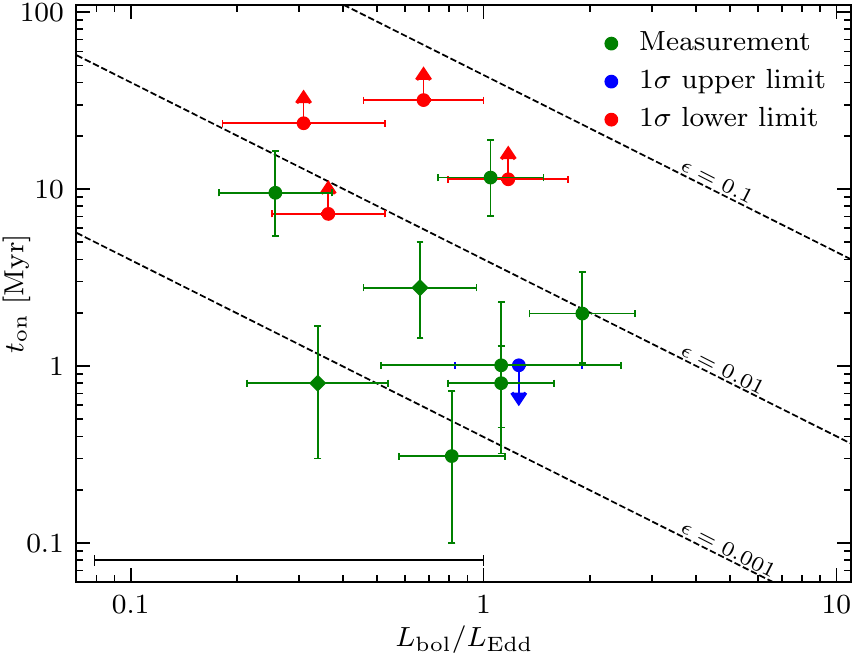}
\caption{\label{fig:he2tqr}
Inferred quasar on-time $t_\mathrm{on}$ as a function of Eddington ratio $L_\mathrm{bol}/L_\mathrm{Edd}$ for
13 \ion{He}{ii}-transparent quasars (Table~\ref{tab:bhmass} excluding HE~2347$-$4342).
Quasar on-time measurements and limits are labelled as in Fig.~\ref{fig:he2tqmag}.
Individual error bars are $1\sigma$ statistical errors, while the bottom bar indicates
the $\pm 0.55$\,dex systematic uncertainty in $L_\mathrm{bol}/L_\mathrm{Edd}$.
The dashed lines show the $e$-folding time-scale of SMBH growth for different radiative efficiencies $\epsilon$.
}
\end{figure}

Figure~\ref{fig:he2tqmbh} shows the quasar on-time as a function of black hole mass for the 13 quasars for which
both quantities are available (Table~\ref{tab:bhmass} excluding HE~2347$-$4342).
In the range of black hole mass spanned by our sample ($M_\mathrm{BH}\simeq 10^9$--$10^{10}\,M_\odot$) both quantities
do not correlate, although the  $\pm 0.55$\,dex systematic uncertainty in the virial black hole masses significantly
contributes to the observed scatter. Figure~\ref{fig:he2tqr} shows the quasar on-time as function of the Eddington
ratio $L_\mathrm{bol}/L_\mathrm{Edd}$. The 13 quasars roughly emit at their Eddington limit considering the systematic
uncertainty induced by the virial black hole mass measurements (Equation~\ref{eq:redd}). The quasar on-time also does
not correlate with the Eddington ratio. The dashed lines in Fig.~\ref{fig:he2tqr} indicate the $e$-folding time-scale
for SMBH growth
\begin{align}
\label{eq:salpetertime}
t_\mathrm{S}&=\frac{c\sigma_\mathrm{T}(1-Y/2)}{4\pi Gm_\mathrm{p}}\frac{\epsilon}{\left(1-\epsilon\right)}\left(\frac{L_\mathrm{bol}}{L_\mathrm{Edd}}\right)^{-1}\\\nonumber
&\simeq 397\,\mathrm{Myr}\frac{\epsilon}{\left(1-\epsilon\right)}\left(\frac{L_\mathrm{bol}}{L_\mathrm{Edd}}\right)^{-1}
\end{align}
for different radiative efficiencies $\epsilon$. For $\epsilon=0.1$ \citep[e.g.][]{yu02,ueda14,shankar20} and
$L_\mathrm{bol}=L_\mathrm{Edd}$ one obtains $t_\mathrm{S}\simeq 44$\,Myr.
This is significantly longer than the on-times $t_\mathrm{on}\lesssim 5$\,Myr we infer for half of our sample.
Such small $t_\mathrm{on}$ values may indicate episodic quasar lifetimes, i.e.\ that their SMBHs grew in short bursts
with $t_\mathrm{Q}\ll t_\mathrm{S}$. Unless most of our observed quasars are radiatively inefficient
\citep[$\epsilon\ll 0.1$, e.g.][]{volonteri15,davies19}, their short on-times imply a small mass growth by $\lesssim 10$
per cent in the current quasar episode. However, the occurrence of lower limits $t_\mathrm{on}\gtrsim 10$\,Myr indicates
that not all quasar episodes are that short, so the \ion{He}{ii} proximity zones may sample a possibly very
broad distribution of quasar lifetimes \citep{khrykin21}. 

\section{Discussion}
\label{sect:discussion}

\subsection{Dependence on the \ion{He}{ii} fraction prior}
\label{sect:he2tqpostlow}

The inferred quasar on-times depend on the assumed prior on the initial \ion{He}{ii} fraction in the surrounding IGM.
Similar to \citetalias{khrykin19}, we adopted a uniform prior $0.01\le x_{\ion{He}{ii},0}\le 1$ to reflect the
considerable uncertainty in the local ionization conditions during \ion{He}{ii} reionization.
The measured large-scale \ion{He}{ii} absorption indicates a low median \ion{He}{ii} fraction of 2--3 per cent at $z\simeq 3.1$
\citep{khrykin16,worseck19} with spatial variations that are consistent with semi-numerical models of a fluctuating
\ion{He}{ii}-ionizing background after \ion{He}{ii} reionization \citep{davies14,davies17}.
However, the parameters of these models are not well constrained by measurements of the \ion{He}{ii} effective optical depth,
such that the spatial distribution of the \ion{He}{ii} fraction is still uncertain.

\begin{figure}
\includegraphics[width=\columnwidth]{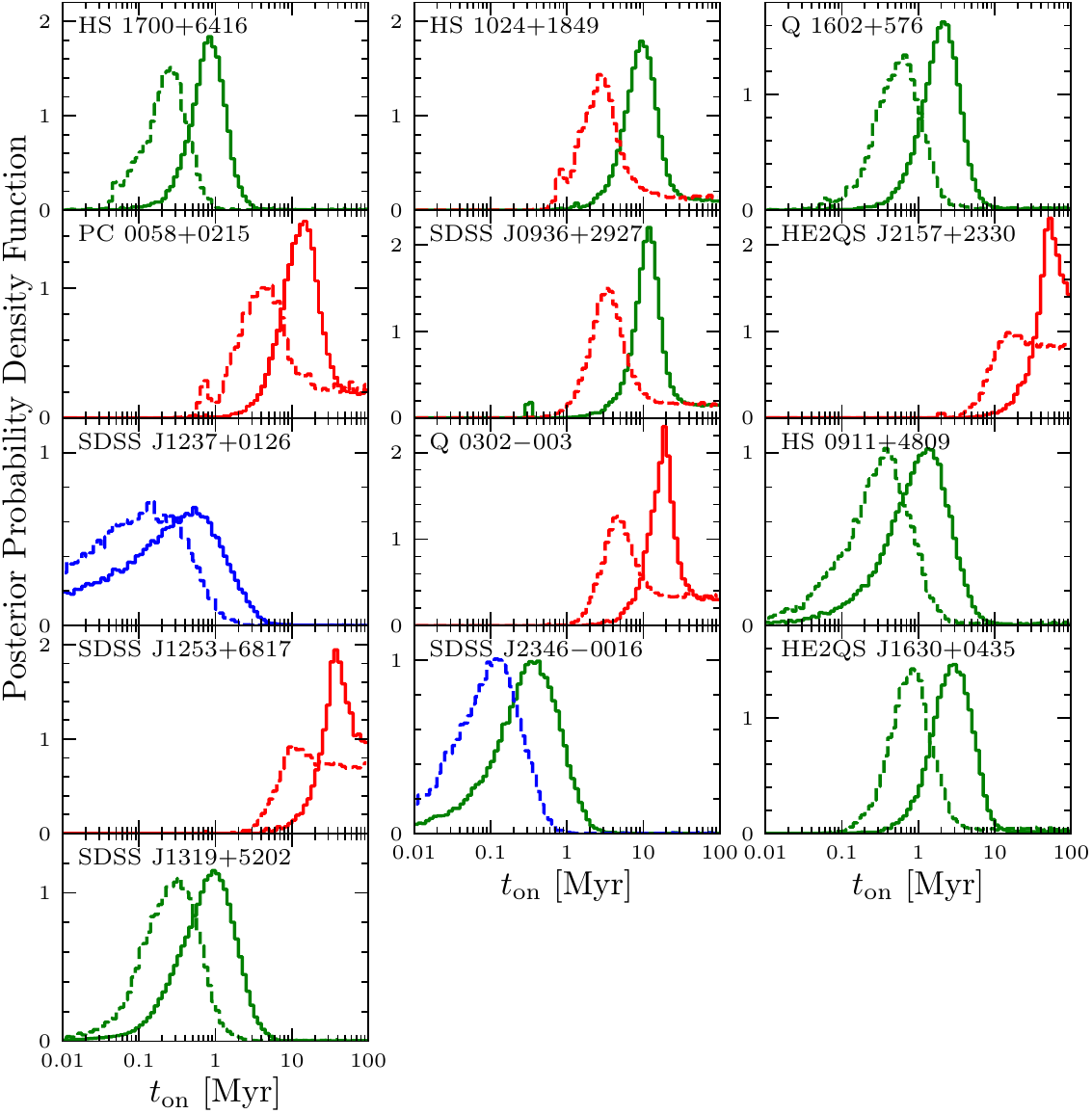}
\caption{\label{fig:he2tqpostlow}
MCMC estimates of the posterior PDFs of the quasar on-time $t_\mathrm{on}$ for two uniform priors
$0.01\le x_{\ion{He}{ii},0}\le 1$ (solid, Fig.~\ref{fig:he2tqpost}) and $0.01\le x_{\ion{He}{ii},0}\le 0.05$ (dashed)
for the 13 quasars with precise systemic redshifts ($\sigma_v\le 400$\,km\,s$^{-1}$).
The colours indicate PDFs resulting in $t_\mathrm{on}$ measurements (green),
$1\sigma$ upper limits (blue), and $1\sigma$ lower limits (red).}
\end{figure}

In order to explore how our results change with the $x_{\ion{He}{ii},0}$ prior, we repeated the analysis with a restricted
uniform prior $0.01\le x_{\ion{He}{ii},0}\le 0.05$ that is consistent with the inferences from the
\ion{He}{ii} Ly$\alpha$ absorption at $2.8<z<3.3$ \citep{worseck19}.
In Figure~\ref{fig:he2tqpostlow} we compare the resulting posterior PDFs for $t_\mathrm{on}$ to the ones from
Section~\ref{sect:ontimes} for the 13 quasars with precisely measured \ion{He}{ii} proximity zone sizes
based on their precise systemic redshifts ($\sigma_v\le 400$\,km\,s$^{-1}$).
Table~\ref{tab:he2prior} compares the on-times for all 22 quasars whose \ion{He}{ii} proximity zone
sizes are not significantly negative, and therefore allow us to constrain $t_\mathrm{on}$
(Table~\ref{tab:sample} excluding HE~2347$-$4342 and HE2QS~J2354$-$2033).
Due to the fewer ionizing photons required to create a large proximity zone, the posterior PDFs shift to lower values,
in agreement with the credible regions for the unrestricted prior at low $x_{\ion{He}{ii},0}$ (Fig.~\ref{fig:kde_mcmc}).
Upper and lower $1\sigma$ limits on $t_\mathrm{on}$ decrease by $\simeq 0.4$\,dex and $\simeq 0.5$\,dex, respectively.
The eight $t_\mathrm{on}$ measurements also shift to lower values by $\simeq 0.5$\,dex, resulting in three more limits
due to the decreasing ratio between the peak of the posterior and its tail at 100\,Myr according to the definition
in Section~\ref{sect:ontimes}. Nevertheless, we still infer a broad range in $t_\mathrm{on}$ from $\lesssim 0.3$ to $\gtrsim 10$\,Myr.
Our constraints can be improved either by more precise systemic redshifts from CO or [\ion{C}{ii}]~158\,$\mu$m emission from
the quasar host galaxies covered at mm to sub-mm wavelengths, or via better priors on $x_{\ion{He}{ii},0}$ from semi-numerical
models of the fluctuating UV background at the end of \ion{He}{ii} reionization \citep{davies17} which is left to future work.

\begin{table}
\caption{\label{tab:he2prior}
On-times $t_\mathrm{on}$ for the 22 quasars considered in this work for two different uniform
priors on the initial IGM \ion{He}{ii} fraction $x_{\ion{He}{ii},0}$.
}
\scriptsize\centering
\begin{tabular}{lcc}
\hline
Quasar	&$0.01\le x_{\ion{He}{ii},0}\le 1$  &$0.01\le x_{\ion{He}{ii},0}\le 0.05$\\
		&$t_\mathrm{on}$ [Myr]              &$t_\mathrm{on}$ [Myr]\\
\hline
HE2QS~J2149$-$0859	&$<0.46$                &$<0.18$\\
HE2QS~J1706$+$5904	&$<0.66$                &$<0.28$\\
SDSS~J1237$+$0126	&$<1.01$                &$<0.36$\\
SDSS~J2346$-$0016	&$0.31^{+0.41}_{-0.21}$ &$<0.22$\\
SDSS~J0818$+$4908	&$<2.42$                &$<0.88$\\
HE2QS~J0916$+$2408  &$<2.77$                &$<1.02$\\
HS~0911$+$4809		&$1.01^{+1.29}_{-0.69}$ &$0.31^{+0.44}_{-0.22}$\\
HE2QS~J0233$-$0149	&$<6.24$                &$<2.36$\\
Q~1602$+$576		&$1.98^{+1.41}_{-0.94}$ &$0.56^{+0.53}_{-0.30}$\\
PC~0058$+$0215		&$>7.24$                &$>2.05$\\
HS~1700$+$6416		&$0.80^{+0.50}_{-0.35}$ &$0.23^{+0.18}_{-0.12}$\\
SDSS~J0936$+$2927	&$11.62^{+7.37}_{-4.57}$&$>2.11$\\
HS~1024$+$1849		&$9.53^{+6.83}_{-4.08}$ &$>1.52$\\
SDSS~J1253$+$6817	&$>23.55$               &$>8.22$\\
Q~0302$-$003		&$>11.36$               &$>3.33$\\
HE2QS~J2157$+$2330	&$>31.84$               &$>10.87$\\
HE2QS~J2311$-$1417	&$<0.86$                &$<0.29$\\
SDSS~J1614$+$4859	&$<5.98$                &$<2.89$\\
SDSS~J1711$+$6052	&$<7.48$                &$<3.66$\\
SDSS~J1319$+$5202	&$0.80^{+0.88}_{-0.50}$ &$0.26^{+0.31}_{-0.16}$\\
SDSS~J1137$+$6237	&$>1.34$                &$>0.71$\\
HE2QS~J1630$+$0435	&$2.77^{+2.27}_{-1.33}$ &$0.82^{+0.83}_{-0.40}$\\
\hline
\end{tabular}
\end{table}

\subsection{Implications for SMBH growth}

\begin{figure}
\includegraphics[width=\columnwidth]{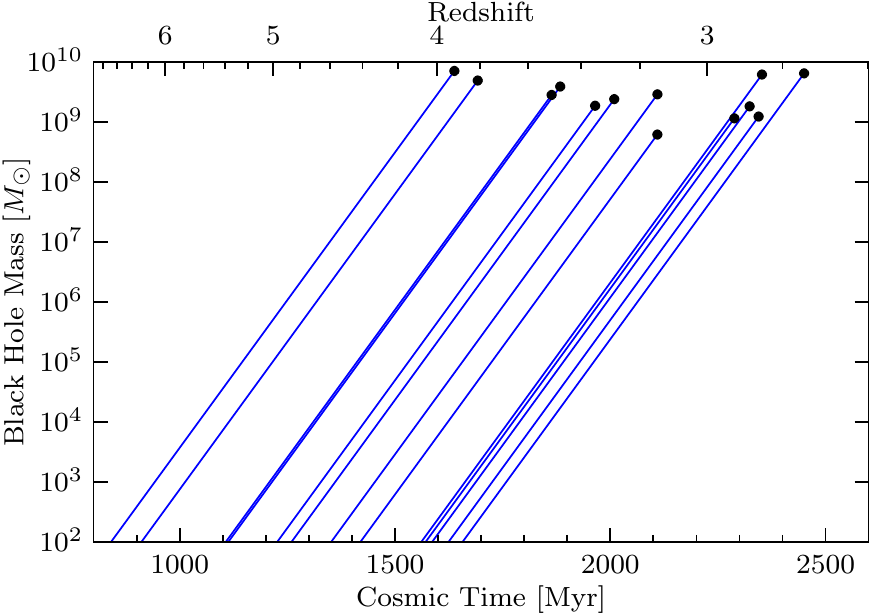}
\caption{\label{fig:he2bhgrowth}
SMBH growth histories of the 13 \ion{He}{ii}-transparent quasars with on-time constraints and measured black hole masses (circles).
The lines show their exponential growth history assuming $\epsilon=0.1$ and a constant Eddington ratio $L_\mathrm{bol}/L_\mathrm{Edd}=1$.
}
\end{figure}

Here we consider the implications of our inferred quasar on-times for the growth histories of their SMBHs to their measured masses.
In our modelling we assumed that each quasar shone at a constant ``light bulb'' luminosity $L_\mathrm{bol}$ for the
time $t_\mathrm{on}$ prior to our observation. Because the inferred on-times are generally much smaller than the Salpeter time,
the light bulb model is still a good approximation to the standard model of exponential mass
growth\footnote{For exponential growth at constant Eddington ratio, Equations~\eqref{eq:redd} and \eqref{eq:bhgrowth}
imply that during $t_\mathrm{on}$ the quasar absolute magnitude increases by $\Delta M_{1450}\simeq 1.19 t_\mathrm{on}/t_\mathrm{S}$,
which is small for most quasars in our sample.}
\begin{equation}
\label{eq:bhgrowth}
M_\mathrm{BH}(t)=M_\mathrm{seed}e^{(t-t_\mathrm{seed})/t_\mathrm{S}}
\end{equation}
for a black hole with a constant radiative efficiency and Eddington ratio that had a seed mass $M_\mathrm{seed}$ at some
cosmic time $t_\mathrm{seed}$. Figure~\ref{fig:he2bhgrowth} illustrates the exponential growth scenario for the 13
\ion{He}{ii}-transparent quasars with on-time constraints and estimated black hole masses.
The growth times $t_\mathrm{gr}=t-t_\mathrm{seed}$ from a stellar remnant seed mass $\sim 100\,M_\odot$ are $t_\mathrm{gr}\sim 700$\,Myr,
and for continuous unobscured SMBH growth these would be equal to the on-times $t_\mathrm{on}$.
For about half of our sample (7 out of 13) the on-times inferred from the \ion{He}{ii} proximity zone sizes
($t_\mathrm{on}\lesssim 5$\,Myr) are considerably shorter than the Salpeter time $t_\mathrm{S}\ll t_\mathrm{gr}$,
implying SMBH growth during episodic quasar activity and/or in obscured phases.

Episodic quasar activity on a wide range of time-scales has been predicted by many models of quasar and black hole co-evolution
\citep[e.g.][]{ciotti01,hopkins06,novak11,angles-alcazar17}. Consider for simplicity a ``blinking light bulb'' model in which
every quasar episode of $t_\mathrm{Q}=20$\,Myr is followed by a quiescent phase of $t_\mathrm{off}=30$\,Myr.
In this case, our observations sample random times $t_\mathrm{on}\le t_\mathrm{Q}$, and after each quasar episode the surrounding
IGM will re-equilibrate to the \ion{He}{ii} fraction implied by the UV background
($t_\mathrm{eq}\approx\Gamma_\ion{He}{ii}^{-1}\simeq$10--30\,Myr at the redshifts of interest), causing the proximity zone to disappear.
If the SMBHs grow exponentially, but not in the quiescent phases, the required time
$t_\mathrm{blink}=t_\mathrm{gr}(t_\mathrm{Q}+t_\mathrm{off})/t_\mathrm{Q}\sim 1750$\,Myr is barely sufficient to explain the SMBH masses
in Fig.~\ref{fig:he2bhgrowth}. Quasars would have to blink over the age of the Universe to acquire their black hole masses.
The problem is exacerbated by shorter quasar lifetimes implied at lower initial \ion{He}{ii} fractions in the end stages of \ion{He}{ii}
reionization (Section~\ref{sect:he2tqpostlow}). This indicates that either the off-times must be shorter than the equilibration time,
such that the proximity zones do not disappear (see \citealt{davies20} for the similar case of $z\sim 6$ \ion{H}{i} proximity zones)
or that black holes continue to grow in obscured phases, i.e.\ during $t_\mathrm{off}$.
X-ray-selected samples of Active Galactic Nuclei (AGN) revealed that the fraction of Compton-thin
(equivalent line-of-sight hydrogen column density $N_\mathrm{H}=10^{22}$--$10^{24}$\,cm$^{-2}$) obscured AGN depends on X-ray luminosity
and redshift \citep[e.g.][]{merloni14,ueda14,buchner15,aird15}. While at $z\sim 2$ the obscured fraction decreases with luminosity from
$\sim 70$ per cent to $\sim 30$ per cent \citep{ueda14,aird15}, at $z\sim 3$ the obscured fraction is $\sim 60$ per cent independent of
luminosity \citep{buchner15,aird15}, possibly increasing further toward higher redshifts \citep{vito18}.
It is therefore plausible that much of the mass growth occurred when the SMBH was obscured by gas and dust, such that SMBH mass and the
duration of the UV-luminous quasar phase are uncorrelated.
We will consider more complex lightcurves than the simple light bulb model in future work.

\section{Conclusions}
\label{sect:conclusions}

We have used a sample of 17 $2.74<z_\mathrm{em}<3.51$ \ion{He}{ii}-transparent quasars with science-grade (S/N$\gtrsim 3$)
\textit{HST}/COS spectra \citep{worseck19} to measure the sizes of their highly ionized \ion{He}{ii} proximity zones.
Given that these zones typically span only a few pMpc, precise measurements are often hampered by quasar redshift error.
Therefore, we obtained ancillary near-infrared spectroscopy to measure accurate and precise systemic redshifts of 12 quasars
from low-ionization UV and optical emission lines (\ion{Mg}{ii}, H$\beta$, [\ion{O}{iii}]) that also allow for estimates
of the quasar black hole masses and Eddington ratios. Together with two $z_\mathrm{em}>3.6$ quasars from \citetalias{khrykin19}
and excluding the peculiar quasar HE~2347$-$4342 \citep[e.g.][]{reimers97}, we have compiled the first statistical sample of
13 quasars with accurate and precise \ion{He}{ii} proximity zone sizes ($\sigma_{R_\mathrm{pz}}\lesssim 1$\,pMpc).
Our main results are the following:

\begin{enumerate}
\item \ion{He}{ii} proximity zone sizes span a large range 2\,pMpc\,$\lesssim R_\mathrm{pz}\lesssim15$\,pMpc.
Nine out of 13 quasars with precise systemic redshifts have $R_\mathrm{pz}>5$\,pMpc, suggesting that large proximity zones
are common, and that quasar redshift error significantly limits further use of the remaining \ion{He}{ii}-transparent quasars. 

\item \ion{He}{ii} proximity zone sizes do not correlate with quasar UV luminosity or redshift (Figs.~\ref{fig:rpzmag} and \ref{fig:rpzzem}).
Given their weak sensitivity to the \ion{He}{ii} fraction in the ambient IGM \citep{khrykin16,khrykin19},
the factor $\sim 5$ spread of $R_\mathrm{pz}$ at similar luminosity is mainly due to variations in the individual quasar on-time,
but variations in the \ion{He}{ii} fraction due to patchy \ion{He}{ii} reionization and IGM density fluctuations contribute as well.

\item Exploiting the sensitivity of $R_\mathrm{pz}$ to quasar on-times $t_\mathrm{on}$ shorter than the equilibration time
of \ion{He}{ii} in the ambient IGM $t_\mathrm{eq}=10$--30\,Myr at these redshifts \citep{khrykin16,worseck19} we have inferred
individual quasar on-times using our Bayesian framework developed in \citetalias{khrykin19}. The 13 quasars with precise
\ion{He}{ii} proximity zone sizes span a large range in on-time from $t_\mathrm{on}\lesssim 1$\,Myr to $t_\mathrm{on}>30$\,Myr,
larger than the typical $\pm 0.3$\,dex statistical uncertainty due to remaining redshift error, the IGM density field,
and the initial \ion{He}{ii} fraction prior to quasar activity. 

\item The quasar on-time neither correlates with quasar luminosity (Fig.~\ref{fig:he2tqmag}), nor with black hole mass
(Fig.~\ref{fig:he2tqmbh}) or Eddington ratio (Fig.~\ref{fig:he2tqr}).
\end{enumerate}

The predominantly short quasar on-times $t_\mathrm{on}<10$\,Myr and the lack of correlation with the black hole properties suggest
that our observations sample the distribution of episodic quasar lifetimes. Unless these quasars are radiatively highly inefficient \citep{davies19},
their black holes must have grown in bursts significantly shorter than the $e$-folding time-scale $t_\mathrm{S}\sim 44$\,Myr.
Such short quasar lifetimes suggest a long quasar duty cycle that is, however, not well constrained given the age of the Universe at $z<4$ ($>1.6$\,Gyr).
This is different to the situation for $z\gtrsim 6$ \ion{H}{i} proximity zones that are insensitive to quasar on-times
$t_\mathrm{on}\gtrsim 0.1$\,Myr \citep{eilers17,eilers18,davies20}, but probe the growth of SMBHs $\lesssim 1$\,Gyr after the Big Bang.
If SMBHs of quasars at $z\sim 6$ accreted similarly to their counterparts at $z\sim 3$, most of their mass must have been built up during
phases of obscuration or radiative inefficiency.

The sensitivity of individual \ion{He}{ii} proximity zones to the time-scale of prior quasar activity of up to $\sim 30$\,Myr offers
a unique opportunity to constrain the underlying distribution of episodic quasar lifetimes \citep{khrykin21},
which can be compared to predictions from models of galaxy and black hole co-evolution. Moreover, we anticipate to more than double
the sample of \ion{He}{ii}-transparent quasars with precise systemic redshifts in our ongoing joint programme with \textit{HST}/COS
and Gemini/GNIRS (PI Worseck) to further resolve quasar activity on time-scales of several tens of Myr.

\section*{Acknowledgements}

We would like to thank Robert Simcoe for sharing his reduced near-infrared spectrum of HE~2347$-$4342 and
Feige Wang for his Python code to fit the near-infrared spectra.
We would also like to thank Anna-Christina Eilers and Frederick Davies for valuable discussions.

GW has been partially supported by the Deutsches Zentrum f\"ur Luft- und Raumfahrt (DLR) grant 50OR1720.
ISK acknowledges support from the grant of the Russian Foundation for Basic Research (RFBR) No. 18-32-00798.

This work is partly based on observations obtained at the Hale Telescope, Palomar Observatory as part of a continuing
collaboration between the California Institute of Technology, NASA Jet Propulsion Laboratory, Yale University,
and the National Astronomical Observatories of China. This material is based upon work supported by the
Association of Universities for Research in Astronomy (AURA) through the National Science Foundation under AURA
Cooperative Agreement AST 0132798 as amended.

The LBT is an international collaboration among institutions in the United States, Italy and Germany.
LBT Corporation partners are: The University of Arizona on behalf of the Arizona Board of Regents; Istituto Nazionale di Astrofisica,
Italy; LBT Beteiligungsgesellschaft, Germany, representing the Max-Planck Society, The Leibniz Institute for Astrophysics Potsdam,
and Heidelberg University; The Ohio State University, and The Research Corporation, on behalf of The University of Notre Dame,
University of Minnesota and University of Virginia.

This work is partly based on observations collected at the European Southern Observatory under ESO programme 290.A-5094.

This work is partly based on observations collected at the Centro Astron\'omico Hispano Alem\'an (CAHA) at Calar Alto,
operated jointly by the Max Planck Institute for Astronomy and the Instituto de Astrof\'isica de Andaluc\'ia (CSIC).

This work is partly based on data obtained from Lick Observatory, owned and operated by the University of California.

Some of the data presented herein were obtained at the W.~M.
Keck Observatory, which is operated as a scientific partnership
among the California Institute of Technology, the University of
California and the National Aeronautics and Space Administration.
The Observatory was made possible by the generous
financial support of the W.~M. Keck Foundation. The authors
wish to recognize and acknowledge the very significant cultural
role and reverence that the summit of Mauna Kea has always
had within the indigenous Hawaiian community. We are most
fortunate to have the opportunity to conduct observations from
this mountain.

This research has made use of the Keck Observatory Archive (KOA), which is operated by the W.~M. Keck Observatory
and the NASA Exoplanet Science Institute (NExScI), under contract with the National Aeronautics and Space Administration.
Some data presented in this work were obtained from the Keck Observatory Database of Ionized Absorbers
toward Quasars (KODIAQ), which was funded through NASA ADAP grant NNX10AE84G.

This paper includes data gathered with the $6.5$\,m Magellan Telescopes located at Las Campanas Observatory, Chile.

Funding for the Sloan Digital Sky Survey IV has been provided by the Alfred P. Sloan Foundation,
the U.S. Department of Energy Office of Science, and the Participating Institutions. SDSS-IV acknowledges
support and resources from the Center for High-Performance Computing at
the University of Utah. The SDSS web site is www.sdss.org.
SDSS-IV is managed by the Astrophysical Research Consortium for the 
Participating Institutions of the SDSS Collaboration including the 
Brazilian Participation Group, the Carnegie Institution for Science, 
Carnegie Mellon University, the Chilean Participation Group, the French Participation Group, Harvard-Smithsonian Center for Astrophysics, 
Instituto de Astrof\'isica de Canarias, The Johns Hopkins University, Kavli Institute for the Physics and Mathematics of the Universe (IPMU) / 
University of Tokyo, the Korean Participation Group, Lawrence Berkeley National Laboratory, 
Leibniz Institut f\"ur Astrophysik Potsdam (AIP),  
Max-Planck-Institut f\"ur Astronomie (MPIA Heidelberg), 
Max-Planck-Institut f\"ur Astrophysik (MPA Garching), 
Max-Planck-Institut f\"ur Extraterrestrische Physik (MPE), 
National Astronomical Observatories of China, New Mexico State University, 
New York University, University of Notre Dame, 
Observat\'ario Nacional / MCTI, The Ohio State University, 
Pennsylvania State University, Shanghai Astronomical Observatory, 
United Kingdom Participation Group,
Universidad Nacional Aut\'onoma de M\'exico, University of Arizona, 
University of Colorado Boulder, University of Oxford, University of Portsmouth, 
University of Utah, University of Virginia, University of Washington, University of Wisconsin, 
Vanderbilt University, and Yale University.

The Pan-STARRS1 Surveys (PS1) and the PS1 public science archive have been made possible through contributions by
the Institute for Astronomy, the University of Hawaii, the Pan-STARRS Project Office, the Max-Planck Society and its
participating institutes, the Max Planck Institute for Astronomy, Heidelberg and the Max Planck Institute for
Extraterrestrial Physics, Garching, The Johns Hopkins University, Durham University, the University of Edinburgh,
the Queen's University Belfast, the Harvard-Smithsonian Center for Astrophysics, the Las Cumbres Observatory Global
Telescope Network Incorporated, the National Central University of Taiwan, the Space Telescope Science Institute,
the National Aeronautics and Space Administration under Grant No. NNX08AR22G issued through the Planetary Science
Division of the NASA Science Mission Directorate, the National Science Foundation Grant No. AST-1238877, the University of Maryland,
Eotvos Lorand University (ELTE), the Los Alamos National Laboratory, and the Gordon and Betty Moore Foundation.

This research made use of \textsc{Astropy} \citep{astropy13,astropy18}, \textsc{Numpy} \citep{numpy11} and \textsc{Matplotlib} \citep{matplotlib07}.

\section*{Data Availability}
The data underlying this article will be shared on reasonable request to the corresponding author.



\bibliographystyle{mnras}
\bibliography{he2prox}



\appendix

\section{Mock \textit{HST}/COS \ion{He}{ii} proximity zone spectra}
\label{sect:mocks}

\begin{figure*}
\includegraphics[width=\textwidth]{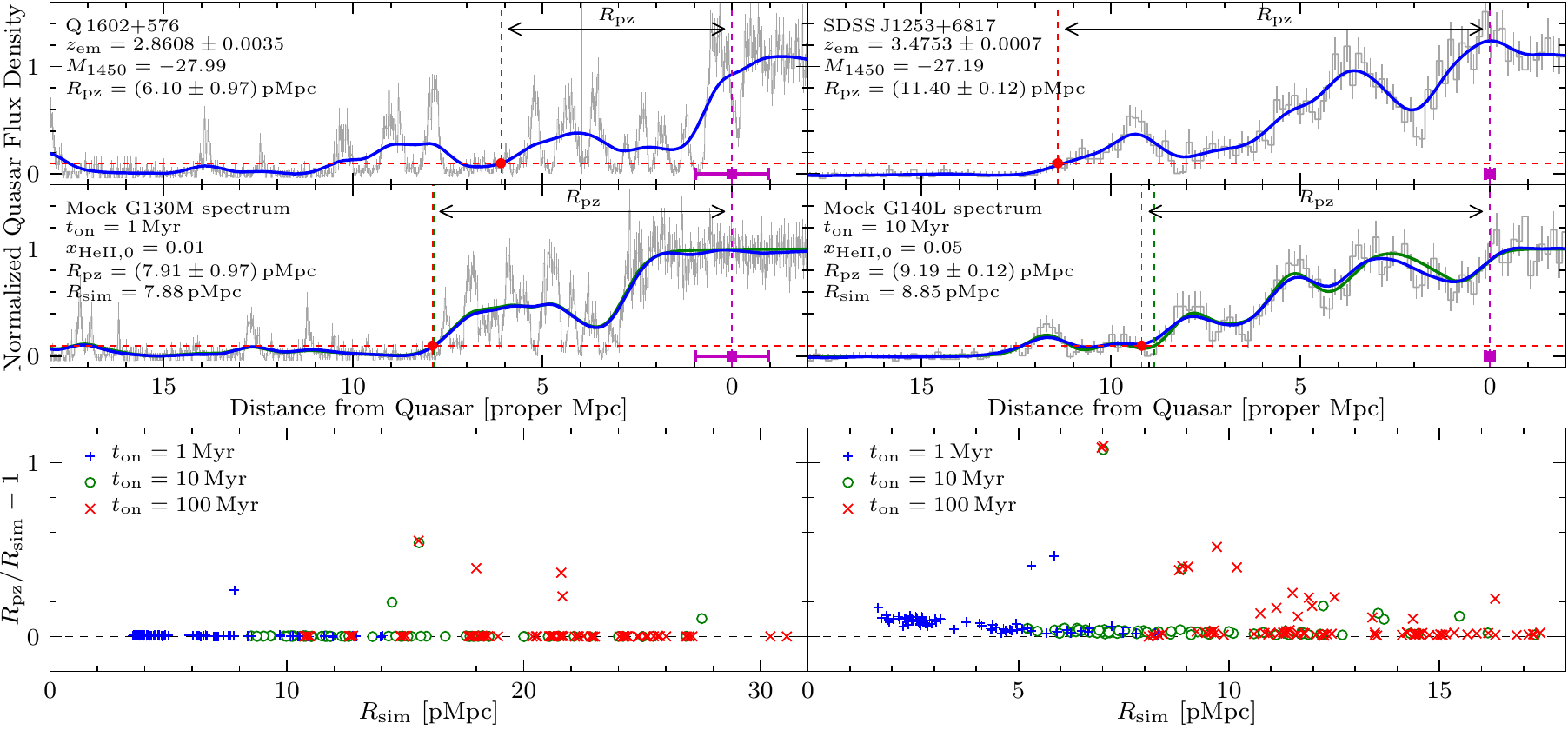}
\caption{\label{fig:he2proxmock}
Comparison of observed \textit{HST}/COS \ion{He}{ii} quasar proximity zone spectra (top panels, labelled)
to representative realistic mock spectra (middle panels).
The left (right) panels show COS G130M (G140L) spectra, plotted in grey with statistical $1\sigma$ Poisson errors.
Distances are for the \ion{He}{ii} Ly$\alpha$ transition relative to the quasar at redshift $z_\mathrm{em}$,
with negative distances indicating pixels in the quasar continuum.
The violet squares with error bars mark the quasar redshift uncertainties. The blue lines show the flux smoothed
with a Gaussian filter with FWHM 1\,pMpc. The red dots mark the measured $R_\mathrm{pz}$.
The green solid lines in the middle panels show the smoothed high-resolution noise-free \ion{He}{ii} transmission
from our radiative transfer models employed in our MCMC analysis, yielding a different proximity zone size
$R_\mathrm{sim}$ (green dashed).
The bottom panels show the relative deviation of $R_\mathrm{pz}$ with respect to $R_\mathrm{sim}$ in the
radiative transfer models for three values of the quasar on-time $t_\mathrm{on}$ (labelled). The dashed lines mark zero deviation.
}
\end{figure*}

In our radiative transfer models, the \ion{He}{ii} proximity zone sizes were determined from high-resolution
($\mathrm{d}r=11.9$ comoving kpc, $\mathrm{d}v=0.86$--$0.93$\,km\,s$^{-1}$ at $z_\mathrm{em}=2.74$--$3.5$)
noise-free \ion{He}{ii} proximity zone Ly$\alpha$ transmission spectra that had not been degraded to the spectral
resolution and quality of the actual \textit{HST}/COS spectra. We explored the consequences of this simplification
with fully forward-modelled mock \textit{HST}/COS \ion{He}{ii} proximity zone spectra of two quasars from our sample
(Q~1602$+$576 taken with the G130M grating and SDSS~J1253$+$6817 taken with the G140L grating),
analogously to \citet{worseck16,worseck19}.
For twelve different combinations of initial \ion{He}{ii} fraction $x_{\mathrm{HeII},0}\in\lbrace 0.01,0.05,0.5,1.0\rbrace$
and quasar on-time $t_\mathrm{on}\in\lbrace 1,10,100\rbrace$\,Myr twenty model spectra were
convolved with the respective \textit{HST}/COS line-spread functions.
Given the actual quasar continuum flux, grating sensitivity, exposure time, background conditions and spectral binning,
expected COS counts per pixel were computed from the convolved \ion{He}{ii} transmission spectra.
Realistic COS Poisson counts were simulated as Poisson deviates of the expected counts, and then converted back
to \ion{He}{ii} transmission. Finally, $R_\mathrm{pz}$ was determined in the same way as for the observed spectra.
Redshift error was not included here.

Figure~\ref{fig:he2proxmock} shows the observed \textit{HST}/COS \ion{He}{ii} proximity zone spectra and representative
mock spectra of Q~1602$+$576 and SDSS~J1253$+$6817. We chose the combinations of $x_{\mathrm{HeII},0}$ and $t_\mathrm{on}$
the $R_\mathrm{pz}$ distribution of which match best the measured values. Apart from the small-scale structure in
the proximity zone sourced by the density field, the mock spectra resemble the observed spectra very well.
The bottom panels show the relative deviations of the proximity zone sizes determined in the mock spectra ($R_\mathrm{pz}$)
with respect to the ones in the high-resolution noise-free model spectra ($R_\mathrm{sim}$).
For most of the 240 mock spectra per quasar the values are very similar, in particular for the range of quasar on-times
our method is sensitive to ($t_\mathrm{on}<30$\,Myr). For long quasar on-times $t_\mathrm{on}\sim 100$\,Myr, $R_\mathrm{pz}$
is sometimes significantly larger than $R_\mathrm{sim}$ due to subtle deviations between the smoothed \ion{He}{ii}
transmission profiles when accounting for the broad wings of the COS line-spread functions and the coarser binning of the
COS spectra. In G140L spectra, proximity zone sizes $\lesssim 5$\,pMpc are overestimated by up to 10 per cent, primarily due to
the sharp drop of the \ion{He}{ii} transmission profile for short quasar on-times $\lesssim 1$\,Myr.
However, this bias is smaller than the error in $R_\mathrm{pz}$ induced by quasar redshift error (Table~\ref{tab:sample}).
For larger $R_\mathrm{sim}$ the bias monotonically decreases to $\lesssim 2$ per cent, as the \ion{He}{ii} proximity zone
transmission profile drops more gradually. In COS G130M spectra the proximity zone size is generally not significantly
overestimated even at small $R_\mathrm{sim}$, as expected. The intrinsic scatter of $R_\mathrm{pz}$ around $R_\mathrm{sim}$
is small, i.e.\ $R_\mathrm{pz}$ is robustly estimated in the COS spectra used in our analysis (see the middle panels of
Fig.~\ref{fig:he2proxmock}). We verified the robustness of $R_\mathrm{pz}$ in the lowest-quality spectra (S/N$\sim 3$).


\bsp	
\label{lastpage}
\end{document}